\documentclass[
reprint,
amsmath,amssymb,
aps,
prx,
floatfix,
superscriptaddress,
longbibliography
]{revtex4-2}

\usepackage[T1]{fontenc}
\usepackage[utf8]{inputenc}
\usepackage{graphicx}
\usepackage{dcolumn}
\usepackage{bm}
\usepackage{array}
\usepackage{booktabs}
\usepackage{microtype}
\usepackage{mathtools}
\usepackage[ruled,vlined]{algorithm2e}
\usepackage[hidelinks]{hyperref}
\graphicspath{{figs/}}
\newcommand{\assign}{:=}

\newcommand{\cdummy}{\cdot}
\newcommand{\nobracket}{}
\newcommand{\nocomma}{}
\newcommand{\nospace}{}
\newcommand{\tmem}[1]{\emph{#1}}

\newcommand{\tmmathbf}[1]{\bm{#1}}
\newcommand{\tmop}[1]{\operatorname{#1}}
\newcommand{\tmrsub}[1]{_{\mathrm{#1}}}

\newcommand{\python}[1]{\texttt{#1}}
\newcommand{\codestar}[1]{\texttt{#1}}
\newenvironment{itemizedot}{\begin{itemize}}{\end{itemize}}
\newenvironment{tmparmod}[3]{\begin{list}{}{\setlength{\topsep}{0pt}\setlength{\leftmargin}{#1}\setlength{\rightmargin}{#2}\setlength{\parindent}{#3}\setlength{\listparindent}{\parindent}\setlength{\itemindent}{\parindent}\setlength{\parsep}{\parskip}}\item[]}{\end{list}}

\newcounter{tmcounter}

\begin{document}

\title{A Stable, High-Order Time-Stepping Scheme for the Drift-Diffusion Model in Modern Solar Cell Simulation}

\author{Jun DU}
\affiliation{Guangdong Basic Research Center of Excellence for Aggregate Science, School of Science and Engineering, The Chinese University of Hong Kong (Shenzhen), Longgang, Shenzhen, Guangdong 518172, P.\,R.\ China}
\author{Jun YAN}
\email{yanjun@cuhk.edu.cn}
\affiliation{Guangdong Basic Research Center of Excellence for Aggregate Science, School of Science and Engineering, The Chinese University of Hong Kong (Shenzhen), Longgang, Shenzhen, Guangdong 518172, P.\,R.\ China}

\begin{abstract}
This paper presents a one-dimensional transient drift--diffusion simulator for advanced solar cells, integrating a structure-preserving finite-volume spatial discretization with Scharfetter--Gummel--type fluxes and a high-order, L-stable implicit Runge--Kutta (Radau IIA) temporal integrator. The scheme ensures local charge conservation, handles sharp material interfaces, achieves second-order spatial and fifth-order temporal convergence, and exhibits excellent computational efficiency and numerical stability. Its accuracy is verified against the classical depletion approximation in $p$--$n$ junction and validated against four independent software implementations for an organic photovoltaic device under a common protocol. The framework's extensibility is demonstrated by incorporating exciton kinetics in organic solar cells, capturing multi-timescale dynamics, and by modeling mobile ions in perovskite solar cells, reproducing characteristic $\tmem{J}$--$\tmem{V}$ hysteresis without empirical parameters. This work provides a robust, high-order numerical foundation for simulating coupled charge, exciton, and ion transport in next-generation photovoltaic devices.
\end{abstract}

\maketitle

\section{Introduction}

Drift--diffusion (DD) modeling has become a central tool for understanding and
optimizing modern solar cells, based on materials ranging from crystalline
silicon to organic and perovskite semiconductors. By resolving the coupled
evolution of electrostatic potential, electronic carrier densities, and, where
relevant, mobile ionic species at the device scale typically in range of
several hundreds of nanometers, DD simulations bridge microscopic material
parameters (mobilities, recombination coefficients, trap and interfacial state
densities) and macroscopic observables such as current--voltage
characteristics, power conversion efficiency, and transient response.
Foundational textbooks such as Nelson's {\tmem{The Physics of Solar Cells}}
{\cite{nelson2003physics}} have established the DD framework as the standard
continuum description for photovoltaic devices, and it is now routinely used
alongside optoelectronic experiments and commercial TCAD tools. For hybrid
perovskites in particular, transient optoelectronic measurements combined with
DD modeling have been instrumental in confirming the role of ion migration and
field screening in current--voltage hysteresis and slow transients
{\cite{calado2016evidence}}.

Most solar-cell DD models can be understood as variants of the classical van
Roosbroeck system, which couples Poisson's equation for the electrostatic
potential with continuity equations for electrons and holes, supplemented by
constitutive relations for drift--diffusion currents and
recombination--generation processes {\cite{van1950theory}}. This framework has
been rigorously analyzed and extended to include mixed ionic--electronic
transport, non-Boltzmann statistics, and sophisticated interface conditions
{\cite{farrell2017drift}}. From the numerical viewpoint, the van Roosbroeck
system poses several challenges: it is nonlinear and strongly coupled, it
involves small intrinsic length scales such as the Debye length, and in mixed
conductors it contains widely separated time scales associated with electronic
carriers, traps, and mobile ions. Any practical solver must therefore cope
with stiffness while preserving basic physical properties such as local charge
conservation, positivity of carrier densities, and meaningful equilibrium
limits over wide parameter ranges.

A variety of spatial discretization strategies have been developed for the
drift--diffusion equations. Early device simulators often employed
finite-difference schemes on structured grids, while finite-element methods
(FEM) have enabled simulations on complex geometries and in multiphysics
settings (e.g. coupled electro--optical or degradation models). For
one-dimensional or layered photovoltaic structures, however, finite-volume
(FV) discretizations have emerged as particularly attractive because they
enforce local conservation at the level of control volumes and naturally
accommodate material interfaces, steep doping gradients, and discontinuous
coefficients. Structure-preserving FV schemes based on Scharfetter--Gummel
(SG) type fluxes {\cite{1475609}} are widely regarded as the standard for
solving drift-diffusion equations: they are exponentially fitted to the local
electric field, preserve positivity, and recover the correct thermal
equilibrium. Recent work has generalized SG fluxes to degenerate statistics
and multicomponent systems and has established discrete entropy/energy
dissipation properties for FV schemes applied to perovskite solar cell models
{\cite{abdel2021modelling}}. Alternative formulations based on electrochemical
potentials and quasi-Fermi potentials, as in Abdel et al.
{\cite{abdel2021modelling}}, improve conditioning and make thermodynamic
structure more explicit. In parallel, Azzouzi et al. have combined DD
transport with detailed exciton kinetics for organic solar cells
{\cite{azzouzi2022reconciling}}, illustrating how complex exciton energy
transfer process can be incorporated consistently in continuum descriptions.
These studies show that, with suitable discretization schemes, the
drift-diffusion framework can be extended to accommodate more complex device
physics while maintaining stability and accuracy during long-term simulations,
especially under steady-state or equilibrium conditions.

On the software side, several open or semi-open one-dimensional DD codes have
become standard tools in the photovoltaic community. IonMonger, introduced by
Courtier et al., is an open-source MATLAB-based simulator for planar
perovskite solar cells that solves a three-layer (ETL/perovskite/HTL) mixed
ionic--electronic model {\cite{courtier2019ionmonger}}. The governing
equations are discretized in space by a second-order finite-element scheme on
a graded mesh resolving narrow Debye layers, and temporal integration of the
resulting DAE system is delegated to MATLAB's stiff multistep solver
{\codestar{ode15s}}. Importantly, IonMonger explicitly does not use SG-type
fluxes, instead relying on a standard FEM discretization tailored to
Debye-layer resolution, and in its three-layer formulation only the majority
carrier is treated in each transport layer (electrons in the ETL, holes in the
HTL). Minority carriers in these layers are not evolved as independent
variables, which complicates the treatment of nearly intrinsic or weakly doped
transport layers and single-carrier experiments where minority-carrier
dynamics in the transport layers are relevant. IonMonger 2.0 extends and
refactors the original implementation but retains the MATLAB
{\codestar{ode15s}} backbone {\cite{clarke23}}, tying time integration to a
proprietary black-box solver and limiting direct experimentation with
alternative integrators and linear solvers at the PDE level.

Driftfusion, developed by Calado et al., takes a complementary approach: it
provides a flexible MATLAB front-end for general one-dimensional mixed
ionic--electronic devices that allows an arbitrary number of layers and up to
four mobile species {\cite{calado2022driftfusion}}. Spatial and temporal
discretization, however, are handled by MATLAB's built-in {\codestar{PDEPE}}
solver, which implements a finite-difference/method-of-lines scheme for
parabolic--elliptic systems. Driftfusion introduces a discrete interlayer
interface approach in which properties are graded smoothly across
finite-thickness ``interface regions'' between adjacent layers to encode
heterojunction energetics and interface-specific recombination mechanisms.
While convenient for defining interface models and avoiding explicit jump
conditions (e.g. energy level difference), this effectively inserts fictitious
interfacial layers of nonzero thickness. Unless the mesh is carefully
controlled, a non-negligible fraction of grid points can be spent resolving
these artificial regions rather than the physical bulk layers, which reduces
the effective resolution for a given computational cost. As in IonMonger, the
underlying flux discretization is based on generic finite differences supplied
by {\codestar{PDEPE}}, rather than on SG-type exponentially fitted FV fluxes
designed specifically for drift--diffusion transport.

More recently, broader simulation platforms have emerged that embed DD solvers
in multi-scale environments. SolarDesign is an online photovoltaic simulation
and design platform that couples DFT-based material calculations, optical
modeling, DD-based device simulation, and circuit-level analysis within a
unified workflow {\cite{wei2025solardesign}}. For the electrical level,
SolarDesign employs non-uniform SG spatial discretization, semi-implicit time
stepping, and Gummel iteration for the coupled Poisson and current-continuity
equations, and incorporates specialized sub-models for tunneling, exciton
dynamics, and ion migration. SIMsalabim, by Koopmans et al., is another widely
used open-source one-dimensional DD code implemented in Pascal with an
emphasis on speed, scriptability, and extensibility
{\cite{koopmans2022simsalabim}}. It supports steady-state and transient
simulations including illumination, mobile ions, trapping--detrapping, and
dielectric mismatch, and is frequently used for parameter extraction and
model-based data analysis in perovskite and organic solar cells. Both
SolarDesign and SIMsalabim offer rich physics and practical workflows, but
their time integration strategies remain within the class of ``classical''
low-order implicit or semi-implicit schemes; they do not aim at systematically
exploring high-order stiff integrators tailored to the structure of the
semi-discrete DD system.

In addition to the custom solvers developed, several commercial and
open-source simulation tools have been widely used for the modeling of
drift--diffusion processes in organic and perovskite solar cells. SETFOS
(Fluxim) is a commercial suite for optoelectronic device simulation, in which
drift--diffusion transport is coupled with optical field calculations. This
combination enables routine evaluation of current--voltage characteristics,
power-conversion efficiency, and transient responses in tandem devices, and it
has been widely used for device optimization and parameter sensitivity studies
in organic photovoltaics and related systems
{\cite{ruhstaller2008comprehensive}}, {\cite{hausermann2009}}. OghmaNano is a
free simulator that also targets drift--diffusion modeling across a broad
range of semiconductor devices, including organic and perovskite solar cells.
It supports both steady-state and transient calculations and has been employed
in multiple studies of organic thin-film devices as a reference implementation
{\cite{mackenzie2012extracting}}. In practice, such tools are often used in
parallel with in-house solvers for benchmarking, calibration, and
cross-validation.

Despite this substantial progress, most DD simulators for solar cells share
two numerical features. First, aside from a few mathematically oriented FV
schemes, the majority of production codes rely either on generic finite
differences/FEM with fluxes not specifically designed for drift--diffusion, or
on SG schemes coupled mainly to first- or second-order temporal
discretizations. Second, time integration is almost always delegated to
low-order implicit multistep schemes: backward Euler, low-order backward
differentiation formulae (BDF), or general stiff solvers such as
{\codestar{ode15s}} and {\codestar{PDEPE}} that internally use variable-order
BDF/NDF methods {\cite{shampine1997matlab}}. These methods are robust and easy
to use, but long transient simulations (e.g. hysteresis protocols, impedance
spectroscopy, or light soaking over many decades in time) often require very
small time steps to achieve acceptable accuracy, because the global order of
accuracy is low and stringent tolerances are needed when each step entails a
large nonlinear solve. High-order implicit Runge--Kutta (IRK) schemes, and in
particular the Radau IIA family, are a natural candidate to address this
limitation. Radau IIA methods are algebraically stable, L-stable, and stiffly
accurate, making them especially well suited to parabolic and
reaction--diffusion type problems where fast modes must be strongly damped
{\cite{hairer1999stiff}}. Their higher order convergence (e.g. fifth order for
the three-stage Radau IIA scheme) allows a given accuracy to be attained with
significantly fewer, larger time steps than first-order backward Euler or
low-order BDF, at the price of solving a coupled stage system at each step.
Although IRK methods are classical in the ODE/DAE literature and have been
used in some fluid and structural dynamics applications, there are few
systematic applications of high-order IRK combined with structure-preserving
FV/SG discretizations in the context of drift--diffusion modeling for solar
cells.

\begin{figure}[t]
\includegraphics[width=\columnwidth]{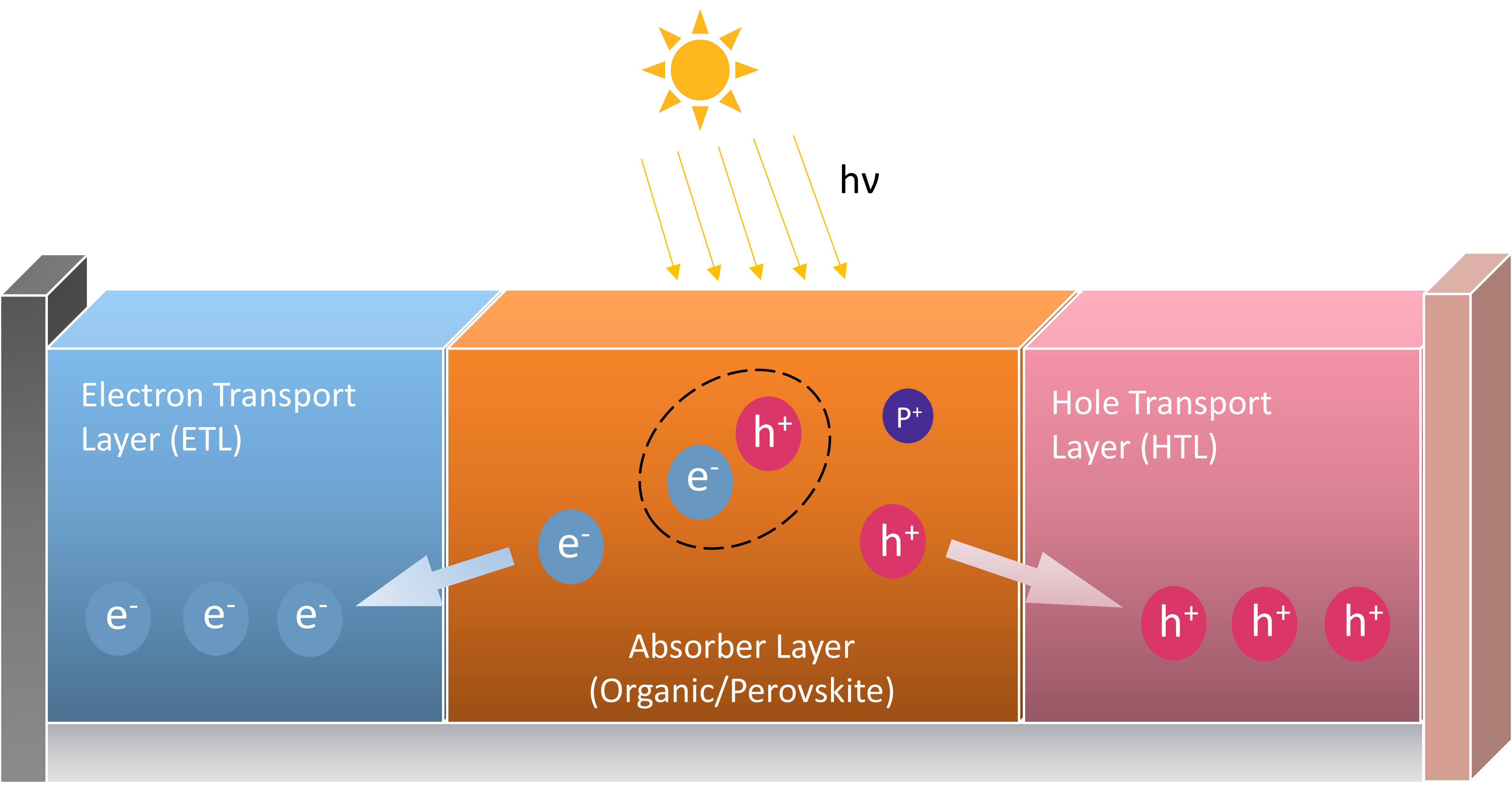}
\caption{Schematic of the multi-layered device structure. The electron-transport layer (ETL), donor/acceptor active layer, and hole-transport layer (HTL) are sandwiched between the cathode and anode. The circled electron-hole pair denotes a photogenerated exciton.}
\label{device}
\end{figure}

In this work we address this gap by developing a one-dimensional
drift--diffusion solar-cell simulator based on a finite-volume discretization
with Scharfetter--Gummel-type fluxes for the Poisson and continuity equations,
combined with an implicit fifth-order Radau IIA scheme for time integration of
the resulting stiff DAE system. Figure \ref{device} illustrates the generic
sandwich-type device geometry (ETL/absorber/HTL) considered in this work. The
finite-volume formulation ensures local charge conservation, handles
heterogeneous multilayer stacks and sharp material interfaces in a natural
way, and is compatible with both density-based and potential-based
formulations, including mobile ionic species. Building on insights from
potential-based and entropy-stable FV schemes {\cite{abdel2021modelling}}, our
spatial discretization is designed to maintain positivity and asymptotic
consistency across strongly inhomogeneous devices. The implicit fifth-order
Radau IIA integrator then enables accurate resolution of transients spanning
many orders of magnitude in time while keeping the number of nonlinear
solutions manageable. In contrast to MATLAB-based frameworks that treat time
stepping as a black box, we integrate the IRK stages directly into a tailored
nonlinear solution strategy, allowing the reuse of Jacobians and
preconditioners across stages and steps and facilitating tight coupling to the
underlying FV structure. Taken together, this space--time discretization is
intended to provide a numerically stable, high-order, and independent
framework capable of capturing both steady-state and transient behavior of
advanced solar-cell architectures, while addressing several of the modeling
and algorithmic limitations identified in existing DD simulators.

\section{Drift Diffusion model}

\subsection{Basic equations}

The drift-diffusion models (Van Roosbroeck's system {\cite{van1950theory}})
are a set of nonlinear equations that comprise continuity equations for
electric charge carriers and the Poisson equation for the electrostatic
potential of the semiconductor devices. Considering the one dimensional and
double-carrier case, the governing equations of the drift diffusion models are
as follows:
\begin{eqnarray}
  \frac{\partial^2 \phi}{\partial x^2} & = & - \frac{q}{\varepsilon} (p - n +
  N_D - N_A), \\
  \frac{\partial n}{\partial t} & = & - \frac{\partial}{\partial x}  \left(
  \frac{\tmmathbf{J}_n}{- q} \right) + G - R, \\
  \frac{\partial p}{\partial t} & = & - \frac{\partial}{\partial x} \left(
  \frac{\tmmathbf{J}_p}{q} \right) + G - R, 
\end{eqnarray}
carrier drift-diffusion fluxes $\tmmathbf{J}_n, \tmmathbf{J}_p$ can be
expressed by quasi-Fermi levels $E_{\tmop{Fn}}, E_{\tmop{Fp}}$:
\begin{equation}
  \tmmathbf{J}_n = n \mu_n  \frac{\partial E_{\tmop{Fn}}}{\partial x}, \qquad
  \tmmathbf{J}_p = p \mu_p  \frac{\partial E_{\tmop{Fp}}}{\partial x} .
\end{equation}
Here, we set spatial domain in $[0, L] = \Omega = \Omega_1 \cup \Omega_2
\cdots \cup \Omega_K \subset \mathbb{R}$, where $\Omega_1, \cdots \Omega_K$
represent different semiconductor material regions (totally $K$ regions), and
$t \in [t_0, t_f]$ represents time. The relevant physical parameters are as
below:
\begin{itemizedot}
  \item $\phi, n, p$ are unknown variables, represent potential, electron
  density and hole density, respectively. The electric field is defined by
  $E=-\partial\phi/\partial x$, and carrier density
  satisfy Fermi-Dirac distribution $n = N_c F_{1 / 2} (\eta_e)$ and $p = N_v
  F_{1 / 2} (\eta_h)$, where $N_c$, $N_v$ are effective density of state
  (eDOS), $F_{1 / 2} (\eta)$ is 1/2 order Fermi-Dirac
  integral (numerically, we use Blakemore approximation~\cite{blakemore1982approximations}, with $F_{1 / 2}(\eta)=1/(e^{\eta}+\gamma)$ and $\gamma=0.27$)
  \begin{equation}
    F_{1 / 2} (\eta) = \frac{2}{\sqrt{\pi}} \int_0^{\infty}
    \frac{\sqrt{x}}{e^{x - \eta} + 1} d x,
  \end{equation}
  where $\eta_e = (- q \phi + E_{\tmop{Fn}} - E_c) / k_B T$, $\eta_h = (q \phi
  - E_{\tmop{Fp}} + E_v) / k_B T$ with charge per electron $q$, Boltzmann
  constant $k_B$ and temperature $T$, $E_c$ and $E_v$ are conduction band and
  valence band, respectively.
  
  \item $\varepsilon$ the permittivity of a material, is given by $\varepsilon
  = \varepsilon_0 \varepsilon_r$, where $\varepsilon_0$ is the vacuum
  permittivity and $\varepsilon_r$ is the relative permittivity.
  
  \item $N_D$ and $N_A$ denote the donor concentration and acceptor
  concentration.
  
  \item $\mu_n$ and $\mu_p$ are the electron and hole mobility, respectively,
  satisfy the Einstein relation
  \begin{equation}
    D = \mu V_t,
  \end{equation}
  where $D$ is carrier diffusion coefficient, $V_t = \frac{k_B T}{q}$.
  
  \item $G$ is generation rate w.r.t. time and space.
  
  \item $R$ is recombination rate w.r.t. $(n, p)$, time and space, e.g.,
  considering Shockley-Read-Hall (SRH) and bimolecular recombination:
  \begin{equation}
    R = \frac{n p - n_i^2}{\tau_p  (n + n_1) + \tau_n  (p + p_1)} + \beta (n p
    - n_i^2), \label{rec}
  \end{equation}
  where $\tau_n, \tau_p$ are SRH time constants for electrons and holes, $n_i$
  is intrinsic density, $n_1$ and $p_1$ are parameters that define the
  dependence of the recombination rate on the trap level $E_{\tmop{trap}}$,
  $\beta$ is the rate constant of bimolecular recombination.
\end{itemizedot}

\subsection{Boundary conditions and initial conditions}

We set up a mixed boundary condition consisting of ohmic contacts and Schottky
contacts, which are mathematically represented as Dirichlet and Robin boundary
conditions, respectively:
\begin{eqnarray}
  \nobracket \phi |_{x = 0} & = & 0, \\
  \nobracket \phi |_{x = L} & = & V_{\tmop{bi}} - V_{\tmop{app}} (t), \\
  \nobracket \tmmathbf{J}_n |_{x = 0} & = & v_{n, l} (n \nobracket |_{x = 0} -
  n_{0, l}) \\
  \nobracket \tmmathbf{J}_n |_{x = L} & = & - v_{n, r} (n \nobracket |_{x = L}
  - n_{0, r}) \\
  \nobracket \tmmathbf{J}_p |_{x = 0} & = & - v_{p, l} (p \nobracket |_{x = 0}
  - p_{0, l}) \\
  \nobracket \tmmathbf{J}_p |_{x = L} & = & v_{p, r} (p \nobracket |_{x = L} -
  p_{0, r}) 
\end{eqnarray}
where $V_{\tmop{bi}}$ is the build-in potential, $V_{\tmop{app}}$ is the
applied voltage, $v_{n / p, r / l}$ represent boundary electron recombination
velocity on left/right boundary, $n_{0, l / r}$ and $p_{0, l / r}$ are
electron and hole density on left/right boundary at equilibrium, given by
\begin{eqnarray}
  n_{0, l / r} & = & N_c F_{1 / 2} \left( \frac{E_F - E_c}{k_B T} \right), \\
  p_{0, l / r} & = & N_v F_{1 / 2} \left( \frac{E_v - E_F}{k_B T} \right), 
\end{eqnarray}
where $E_F$ is Fermi energy level.

For the system in consistency, we require that the initial guess of $(n, p)$
satisfy charge neutrality condition and equilibrium condition (here we set
equilibrium Fermi level $E_F$ at 0)
\begin{eqnarray}
  p - n + N_D - N_A & = & 0, \\
  E_{\tmop{Fn}} = E_{\tmop{Fp}} & = & 0, 
\end{eqnarray}
and assume no electrical field, i.e. the potential $\phi (x) = 0$, for initial
guess. For transient problem, the initial condition of transient term
$\partial_t y$ (for any unknown variable $y$ regard as $y {}'$) is obtained by
solving system $F (y, y {}', t, x)$ using newton's method, referring to
{\cite{shampine2002solving}}.

\section{Numerical Methods and Algorithms}

\subsection{Spatial mesh}

\begin{figure}[t]
\includegraphics[width=\columnwidth]{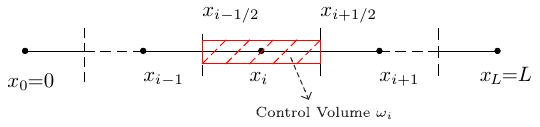}
\caption{A schematic of Spatial Discretization.}
\label{mesh}
\end{figure}

Considering the potential occurrence of large gradients at the boundaries, we
apply the {\tmem{tanh}} grid mapping {\cite{courtier2018fast}} to the
subdomain $\nobracket x \in [0, L_1) \cup (L_1, L_2) \cdots \cup (L_{j - 1},
L_j) \cdots \cup (L_{K - 1} \nobracket, L]$, ensuring good resolution at the
boundaries:
\begin{equation}
\begin{split}
  \frac{x_i - L_{j - 1}}{|L_j - L_{j - 1}|}
  &= \frac{1}{2} \left( \frac{\tanh \left[ \xi \left( \frac{2 i}{\sum_{i = 0}^{N_j} i} - 1 \right) \right]}{\tanh (\xi)} + 1 \right),\\
  &\hspace{7.5em} \tmop{for}\ x_i \in \Omega_j .
\end{split}
\label{tanh}
\end{equation}
where $N_j$ represents node number of the $j$th material, $\xi$ represents the
degree to which points are concentrated near the boundaries (generally let
$\xi = 5$). For convenience, the ghost point $x_0$ and $x_N$ are set to
represent boundary condition, so the mesh consists of $\sum^K_{j = 0} N_j + 2$
nodes. For these subintervals, their control volume are defined as $\omega_i^-
= x_i - x_{i - 1 / 2}$, $\omega_i^+ = x_{i + 1 / 2} - x_i$ and $\omega_i =
\omega_i^+ + \omega_i^-$, as shown in Figure \ref{mesh}.

\subsection{Nondimensionalisation}\label{non-dim}

We introduce dimensionless variables (denoted by a hat) by normalizing each
physical quantity with its characteristic scale:
\begin{align}
  \hat{x} &= \frac{x}{L}, & \hat{t} &= \frac{t}{\tau}, & \hat{\phi} &= \frac{\phi}{V_t}, \notag\\
  \hat{n} &= \frac{n}{N_0}, & \hat{p} &= \frac{p}{N_0}, & \hat{N}_D &= \frac{N_D}{N_0}, \notag\\
  \hat{N}_A &= \frac{N_A}{N_0}, & \hat{\tmmathbf{J}}_n &= \frac{\tmmathbf{J}_n}{q G_0 b}, & \hat{\tmmathbf{J}}_p &= \frac{\tmmathbf{J}_p}{q G_0 b},
  \label{eq:nondimvars}\\
  \hat{G} &= \frac{G}{G_0}, & \hat{R} &= \frac{R}{G_0}, & \hat{v} &= \frac{v N_D}{G_0 b}, \notag\\
  \hat{E}_{\tmop{Fn}} &= \frac{E_{\tmop{Fn}}}{k_B T}, & \hat{E}_{\tmop{Fp}} &= \frac{E_{\tmop{Fp}}}{k_B T}. & & \notag
\end{align}
\begin{align}
  N_0 &= \|N_D, N_A\|_{\max}, & L_D &= \sqrt{\frac{V_t \|\varepsilon\|_{\max}}{q N_0}}, \notag\\
  G_0 &= \frac{N_0 \cdot \|D_n, D_p\|_{\max}}{L^2}, & \tau &= \frac{N_0}{G_0}.
  \label{eq:charscales}
\end{align}
\begin{equation}
  \lambda = \frac{L_D}{L}, \qquad \kappa_n = \frac{D_n N_0}{G_0 L^2}, \qquad \kappa_p = \frac{D_p N_0}{G_0 L^2}.
  \label{eq:nondimparams}
\end{equation}
where $L_D$ is so-called Debye length. From these scales we obtain the
dimensionless parameters
then (after dropping the hats on the parameters and variables) equations can
be rewritten as
\begin{alignat}{2}
\label{poisson}
0 &= -\lambda^2 \frac{\partial \mathbf{E}}{\partial x} + p - n + N_D - N_A,
&\hspace{0.6em}\mathbf{E} &= -\frac{\partial \phi}{\partial x},\\
\label{contin}
\frac{\partial n}{\partial t} &= \frac{\partial \mathbf{J}_n}{\partial x} + G - R,
&\hspace{0.6em}\mathbf{J}_n &= \kappa_n n \frac{\partial E_{\mathrm{Fn}}}{\partial x},\\
\label{contip}
\frac{\partial p}{\partial t} &= -\frac{\partial \mathbf{J}_p}{\partial x} + G - R,
&\hspace{0.6em}\mathbf{J}_p &= \kappa_p p \frac{\partial E_{\mathrm{Fp}}}{\partial x}.
\end{alignat}
Notice that the Debye length is much smaller than the physical characteristic
length, there is a small parameter $\lambda^2$ in front of the electric
potential term in the Poisson equation Eq.(\ref{poisson}), which will result
in a {\tmem{singular perturbation}} problem {\cite{singer2008singular}},
{\cite{he2016energy}}, thus the {\tmem{tanh}} grid mapping Eq.(\ref{tanh}) is
necessary.

\subsection{Finite volume scheme}\label{fvm}

The numerical method consists of five main steps: (1) partitioning the device
into cells with a refined mesh near boundaries and interfaces; (2) integrating
Poisson's equation over each cell; (3) computing electron and hole currents at
the cell faces using the Scharfetter--Gummel formula; (4) combining these face
currents with generation and recombination terms to construct a mass-matrix
differential--algebraic system; and (5) advancing this system with an adaptive
three-stage Radau IIA method.

The same face current is shared by two neighbouring cells. Therefore, current
leaving one cell enters the next cell, which gives local conservation. The same five steps are also used when exciton or ionic variables are added.
Importantly, the time-dependent equations constitute an initial-value problem
(IVP). Each implicit Radau step requires solving nonlinear stage equations,
because the flux terms, Poisson field, recombination, and boundary conditions
depend on the unknown stage state. For steady-state calculations, we set
$\partial n/\partial t=\partial p/\partial t=0$ and solve the resulting
nonlinear algebraic system using Newton's method. Although both formulations are
nonlinear, the transient problem additionally involves time integration of the
IVP and stiff coupling between stages. The main contribution of this work is the
stable, high-order Radau IIA time-stepping scheme.

The general form of time dependent nonlinear transport-reaction problems with
transient term $\partial_t u$, flux term $\tmmathbf{F}$, reaction term $r (u)$
and source term $f$:
\begin{equation}
  \partial_t u = \nabla \cdummy \tmmathbf{F}+ r (u) + f,
\end{equation}
where $u$ is unknown function. Apply the Gauss theorem for the divergence
term, we have
\begin{equation}
  \int_{\Omega} \nabla \cdummy \tmmathbf{F}d \Omega = \oint_{\partial \Omega}
  \tmmathbf{F} \cdummy d\tmmathbf{n}= \sum_k \tmmathbf{F}_{i k} d_{i k},
\end{equation}
where $\partial \Omega$ is the boundary, $\tmmathbf{n}$ is direction vector,
$\tmmathbf{F}_{i k}$ is the midpoint flow density on the edge$(i, k)$whose
direction is from vertex $i$ to vertex $k$, $d_{i k}$ is the length of the
side perpendicular to the opposite side $(i, k)$ on the polygon $\Omega$. In
one dimension case, the transport equation has the form
\begin{equation}
  \int \frac{\partial s (u)}{\partial t} d \omega_i = \sum_k \tmmathbf{F}_{i
  k} d_{i k} + \int (r (u) + f) d \omega_i,
\end{equation}
the integral of the flux divergence at the boundary of the control volume is
\begin{equation}
  \sum_k \tmmathbf{F}_{i k} d_{i k} =\tmmathbf{F}_{i + 1 / 2} -\tmmathbf{F}_{i
  - 1 / 2} .
\end{equation}
For the source and reaction terms we use a piecewise-linear reconstruction and
Simpson quadrature{\cite{cartwright2017simpson}}. For each control volume this
gives:
\begin{equation}
  \omega_i \frac{\partial s (u)}{\partial t} =\tmmathbf{F}_{i + 1 / 2}
  -\tmmathbf{F}_{i - 1 / 2} +\mathcal{S}_i  [r (u)] + \omega_i f_i,
\end{equation}
where an operator $\mathcal{S}_i$ is defined as $\mathcal{S}_i [r (u)] : =
\frac{\omega_i}{6}  [r (u_{i - 1 / 2}) + 4 r (u_i) + r (u_{i + 1 / 2})]$,
furthermore, by using average approximation $r (u_{i + 1 / 2}) \approx \frac{r
(u_i) + r (u_{i + 1})}{2}$ of reaction term (recombinations Eq.(\ref{rec}) are
not excessing 2 order), then Simpson integral operator turns to $\mathcal{S}_i
[r (u)] = \left[ \frac{1}{12} r (u_{i - 1}) + \frac{5}{6} r (u_i) +
\frac{1}{12} r (u_{i + 1}) \right] \omega_i$.

The finite-volume method (FVM) therefore enforces conservation locally. We
write the resulting face differences and averages for a generic variable $u$
as
\begin{eqnarray}
  D_x u_{i + 1 / 2} & : = &  \frac{u_{i + 1} - u_i}{\Delta_{i + 1 / 2}}, \\
  A_x u_{i + 1 / 2}  & : = &  \frac{u_{i + 1} + u_i}{2}, 
\end{eqnarray}
where $\Delta_{i + 1 / 2} = x_{i + 1} - x_i$. For electrical field, we use
finite difference approximation $\tmmathbf{E}_{i + 1 / 2} \approx - D_x
\phi_{i + 1 / 2}$; for electrons/holes fluxes, we employ the Scharfetter
Gummel {\cite{1475609}} scheme (also known as Il'In {\cite{il1969difference}},
Allen-Southwell {\cite{de1955relaxation}}) which is obtained by solving a
local boundary value problem (BVP) across the intervals or can be explained by
a kind of general harmonic average, e.g., by defining $\gamma_e = \ln
\frac{F_{1 / 2} (\eta_e)}{\exp (\eta_e)} - E_c + \ln N_c$, we have $n
\frac{\partial E_{\tmop{Fn}}}{\partial x} = \exp (\phi + \gamma_e) 
\frac{\partial \exp (E_{\tmop{Fn}})}{\partial x}$, which leads to
$\tmmathbf{J}_{n, i + 1 / 2} \approx A_x \kappa_{n, i + 1 / 2}  \frac{\gamma
{}'_{i + 1} - \gamma {}'_i}{\int_{\gamma {}'_i}^{\gamma {}'_{i + 1}} \exp
(\gamma {}') d x}  \frac{\exp (E_{\tmop{Fn}, i + 1}) - \exp (E_{\tmop{Fn},
i})}{\Delta_{i + 1 / 2}}$. Assuming $\gamma {}' = \phi + \gamma_e$ changes
linearly within the subinterval $[x_i, x_{i + 1}]$, one can obtain
\begin{equation}
  \tmmathbf{J}_{n, i + 1 / 2} = \frac{A_x \kappa_{n, i + 1 / 2}}{\Delta_{i + 1
  / 2}} [B (- z_n) n_{i + 1} - B (z_n) n_i],
\end{equation}
where $z_n = (E_{\tmop{Fn}, i + 1} - E_{\tmop{Fn}, i}) - \ln \frac{F_{1 / 2}
(\eta_{e, i + 1})}{F_{1 / 2} (\eta_{e, i})} - \ln \left( \frac{N_{c, i +
1}}{N_{c, i}} \right)$, and $B (x) = \frac{x}{e^x - 1}$ is Bernoulli function.
Similarly, for hole fluxes,
\begin{equation}
  \tmmathbf{J}_{p, i + 1 / 2} = \frac{- A_x \kappa_{p, i + 1 / 2}}{\Delta_{i +
  1 / 2}} [B (z_p) p_{i + 1} - B (- z_p) p_i],
\end{equation}
where $z_p = - (E_{\tmop{Fp}, i + 1} - E_{\tmop{Fp}, i}) - \ln \frac{F_{1 / 2}
(\eta_{h, i + 1})}{F_{1 / 2} (\eta_{h, i})} - \ln \frac{N_{v, i + 1}}{N_{v,
i}}$.

After inserting the face fluxes and boundary conditions, the semi-discrete
equations are a DAE in time. The algebraic rows are Poisson's equation and the
contact relations; the differential rows are the carrier balances:
\begin{widetext}
\begin{eqnarray}
  0 & = & \phi_0,  \label{starDAE}\\
  0 & = & - \lambda^2 (\tmmathbf{E}_{i + 1 / 2} -\tmmathbf{E}_{i - 1 / 2}) +
  \omega_i (p_i - n_i + N_{D, i} - N_{A, i}) \\
  &  & \tmop{for} i = 1, 2, \cdots N - 1, \nonumber\\
  0 & = & \phi_N + V_{\tmop{app}} (t) - V_{\tmop{bi}}, \\
  &  &  \nonumber\\
  0 & = & \tmmathbf{J}_{n, 1 / 2} - v_{n, l} (n_0 - n_{0, l}) \\
  \omega_i  \frac{d n_i}{d t} & = & \tmmathbf{J}_{n, i + 1 / 2}
  -\tmmathbf{J}_{n, i - 1 / 2} + \omega_i G_i -\mathcal{S}_i R \\
  &  & \tmop{for} i = 1, 2, \cdots, N - 1, \nonumber\\
  0 & = & -\tmmathbf{J}_{n, N - 1 / 2} - v_{n, r} (n_N - n_{0, r}), \\
  &  &  \nonumber\\
  0 & = & -\tmmathbf{J}_{p, 1 / 2} - v_{p, l} (p_0 - p_{0, l}), \\
  \omega_i \frac{d p_i}{d t} & = & - (\tmmathbf{J}_{p, i + 1 / 2}
  -\tmmathbf{J}_{p, i - 1 / 2}) + \omega_i G_i -\mathcal{S}_i R \\
  &  & \tmop{for} i = 1, 2, \cdots, N - 1, \nonumber\\
  0 & = & \tmmathbf{J}_{p, N - 1 / 2} - v_{p, r} (p_N - p_{0, r}) . 
  \label{endDAE}
\end{eqnarray}
\end{widetext}

\subsection{Dynamics in Solar Cells}

Beyond purely electronic transport, many modern solar-cell architectures
involve additional internal degrees of freedom such as excitonic populations
in organic blends or mobile ionic defects in metal--halide perovskites. In the
present framework, these processes are treated in a way that is fully
consistent with the nondimensionalization and discretization introduced in
Sections \ref{non-dim} and \ref{fvm}. For organic solar cells, the local LE/CT
exciton kinetics are described by zero-dimensional rate equations at each
spatial grid point, see Eqs.(\ref{le})--(\ref{ct}), which enter the
discretized system as stiff reaction terms coupled to the electronic carrier
densities $n$ and $p$ {\cite{azzouzi2022reconciling}}. For perovskite solar
cells, mobile ionic vacancies obey a one-dimensional conservation law with
drift--diffusion flux $\tmmathbf{J}_P$, so that Eq.(\ref{pscJ}) which is
discretized by the same finite-volume machinery and SG-type fluxes as the
electronic currents. After rescaling with the global characteristic length and
time scales, all additional variables (excitons, ions) simply augment the
semi-discrete DAE system in Eqs.(\ref{starDAE})--(\ref{endDAE}) and are
advanced in time by the common fifth-order Radau IIA integrator. This
highlights that the proposed numerical framework is not tied to a specific
device physics, but can accommodate new dynamic sub-models in a modular way.

\subsubsection{Kinetics of excitons}

For organic solar cells (OSC), we mainly consider the local exciton (LE) state
and the interfacial charge transfer (CT) state using zero-dimensional rate
equations, described by {\cite{azzouzi2022reconciling}}:
\begin{widetext}
\begin{eqnarray}
  \frac{\partial [\tmop{LE}]}{\partial t} & = & G_{\tmop{opt}} -
  k_{\tmop{trans}}^{\tmop{LE} \rightarrow \tmop{CT}} [\tmop{LE}] +
  k_{\tmop{trans}}^{\tmop{CT} \rightarrow \tmop{LE}} [\tmop{CT}] -
  k_{\tmop{rec}}^{\tmop{LE}}  ([\tmop{LE}] - [\tmop{LE}]_0),  \label{le}\\
  \frac{\partial [\tmop{CT}]}{\partial t} & = & k_{\tmop{trans}}^{\tmop{LE}
  \rightarrow \tmop{CT}} [\tmop{LE}] - k_{\tmop{diss}}^{\tmop{CT} \rightarrow
  \tmop{CS}} [\tmop{CT}] - k_{\tmop{trans}}^{\tmop{CT} \rightarrow \tmop{LE}}
  [\tmop{CT}] - k_{\tmop{rec}}^{\tmop{CT}} ([\tmop{CT}] - [\tmop{CT}]_0) +
  B_{\tmop{for}}^{\tmop{CS} \rightarrow \tmop{CT}} n p,  \label{ct}
\end{eqnarray}
\end{widetext}
where $G_{\tmop{opt}}$ is the generation rate obtained from optical model,
$k_{\tmop{trans}}$ is the constant transfer rate between LE and CT states,
$k_{\tmop{diss}}$ is constant dissociation rate that CT state contribute the
free charge carriers, $k_{\tmop{rec}}$ the constant rate for LE and CT state
recombination, $B_{\tmop{for}}$ the constant rate of formation of a CT state
from free charge carriers, $[\tmop{LE}]_0 / [\tmop{CT}]_0$ is the equilibrium
exciton density for the lowest LE and CT state, respectively. Coupled with
drift-diffusion model, the continuity equations of free electrons and holes
turns to describe charge separated (CS) state:
\begin{eqnarray}
  \frac{\partial n}{\partial t} & = & - \frac{\partial}{\partial x} \left(
  \frac{\tmmathbf{J}_n}{- q} \right) + k_{\tmop{diss}}^{\tmop{CT} \rightarrow
  \tmop{CS}}  [\tmop{CT}] - B_{\tmop{for}}^{\tmop{CS} \rightarrow \tmop{CT}} n
  p, \\
  \frac{\partial p}{\partial t} & = & - \frac{\partial}{\partial x}  \left(
  \frac{\tmmathbf{J}_p}{q} \right) + k_{\tmop{diss}}^{\tmop{CT} \rightarrow
  \tmop{CS}}  [\tmop{CT}] - B_{\tmop{for}}^{\tmop{CS} \rightarrow \tmop{CT}} n
  p. 
\end{eqnarray}

\begin{figure}[t]
\includegraphics[width=\columnwidth]{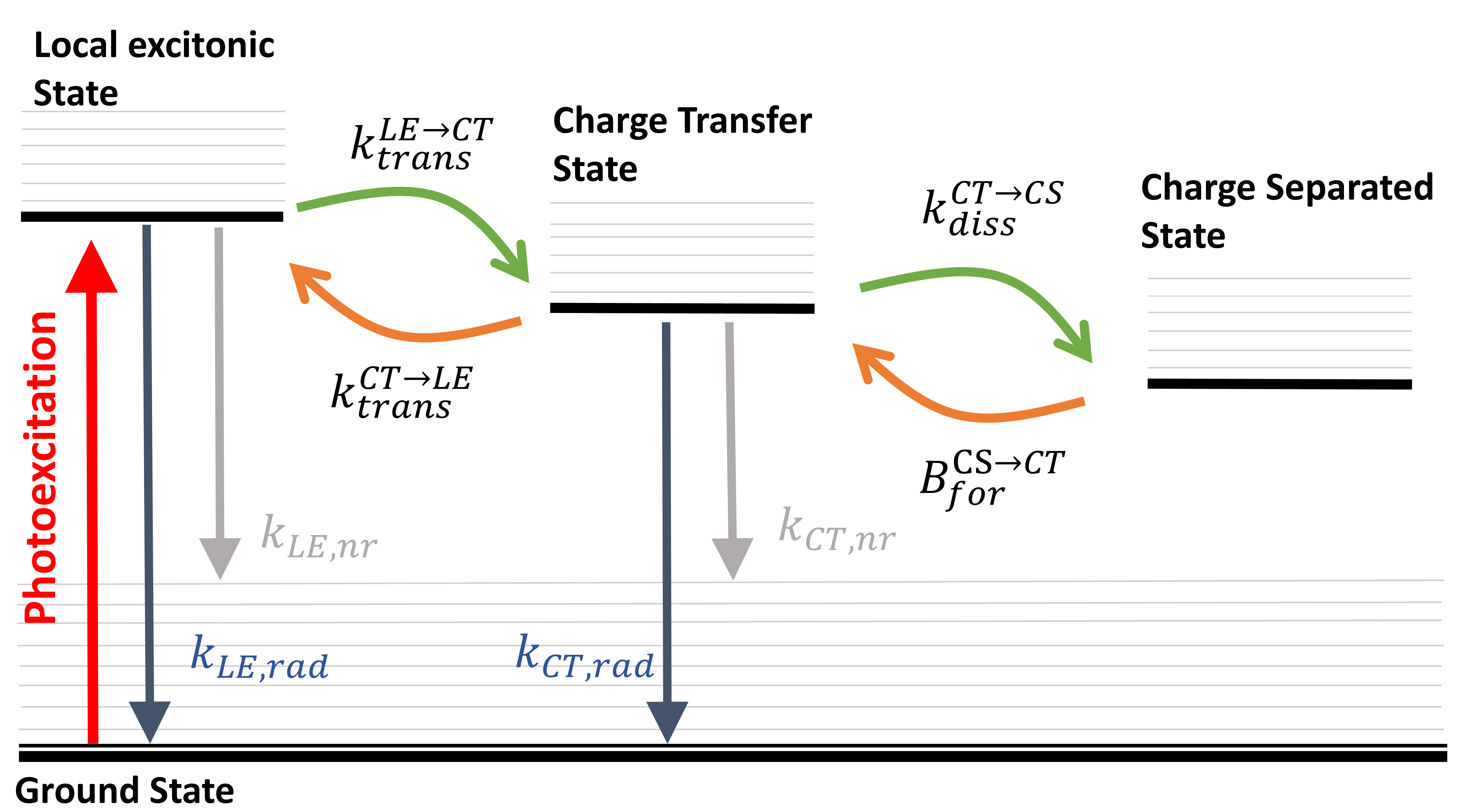}
\caption{Kinetic scheme used for the coupled exciton-charge dynamics. The arrows show photoexcitation, charge transfer, recombination, and free-carrier generation.}
\label{sd}
\end{figure}

Figure \ref{sd} schematically shows the interconversion between LE and CT
states, CT dissociation to the charge-separated (CS) state and the reverse
back-formation, together with radiative/non-radiative loss channels and
photoexcitation (pump). For each excitonic state, the total recombination rate
is the sum of the radiative and non-radiative rates,
$k_{\tmop{rec}}=k_{\tmop{rad}}+k_{\tmop{nr}}$.

\subsubsection{Dynamics of mobile ionic vacancies}

For perovskite solar cells (PSC), the conservation of positively-charged anion
vacancies $P$ takes the form {\cite{courtier2018fast}}:
\begin{equation}
  \frac{\partial P}{\partial t} = - \frac{\partial \tmmathbf{J}_P}{\partial
  x}, \label{pscJ}
\end{equation}
with drift and diffusion current due to ionic motion:
\[ \tmmathbf{J}_P = - D_+  \left( \frac{\partial P}{\partial x} +
   \frac{P}{V_t}  \frac{\partial \phi}{\partial x} \right), \]
where $D_+$ represents anion vacancy diffusion coefficient.

Poisson's equation in perovskite layer turns to:
\[ \frac{\partial^2 \phi}{\partial x^2} = - \frac{q}{\varepsilon} (P - N_I + p
   - n), \]
where $N_I$ is the uniform density of cation vacancies (for satisfying the
global charge neutrality).

\subsection{Time stepping}

The assembled problem has the compact form
\begin{equation}
  M \frac{d y}{d t} = f (t, y),
\end{equation}
where $M$ is the mass matrix, $y=[\phi,n,p]$ is the state vector (augmented by
ions or excitons when present), and $f$ is the assembled residual. The small
Debye length and fast reaction rates make this system stiff, so an L-stable
integrator is required to damp unresolved fast modes without imposing an
artificial restriction on the slow physical dynamics. We use the three-stage
Radau IIA method: it is fifth-order accurate, stiffly accurate, and L-stable.
The implementation adapts the {\python{Radau}} algorithm
{\cite{2020SciPy-NMeth, hairer2006numerical}} to the mass-matrix DAE and exposes
the stage residual and linear solves rather than treating time integration as
a black box.

For an ordinary differential equation (ODE) problem $\frac{d y}{d t} = f (t,
y)$, the updated formula from a time step $t_n$ to $t_{n + 1} = t_n + h$ is
{\cite{butcher1976implementation}}
\begin{eqnarray}
  y_{n + 1} & = & y_n + h \sum_{i = 1}^s b_i K_i, \\
  K_i & = & f (t_n + c_i h, Y_i), i = 1, 2, \cdots, s 
\end{eqnarray}
where stage $i$ is evaluated at $t_n+c_i h$ and
$Y_i=y_n+h\sum_{j=1}^s a_{ij}K_j$. The coefficients $c_i$ set the stage
times, $a_{ij}$ couple the stage solutions, and $b_i$ combine the stages to
obtain $y_{n+1}$. An $s$-stage Radau IIA method has order $2s-1$; therefore,
the three-stage method used here has order five {\cite{wanner1996solving}}.
Table~\ref{tab:radau-coefficients} lists the coefficients with their roles.

\begin{table*}[t]
\caption{Coefficients used by the three-stage Radau IIA method. Rows 1--3
define the internal stages; the last row gives the final update weights.}
\label{tab:radau-coefficients}
\begin{ruledtabular}
\begin{tabular}{ccllll}
Role & Stage & Time $c_i$ & Weight $a_{i1}$ & Weight $a_{i2}$ & Weight $a_{i3}$ \\
\colrule
Internal stage & 1 & $\frac{4-\sqrt{6}}{10}$ & $\frac{88-7\sqrt{6}}{360}$ & $\frac{296-169\sqrt{6}}{1800}$ & $\frac{-2+3\sqrt{6}}{225}$ \\
Internal stage & 2 & $\frac{4+\sqrt{6}}{10}$ & $\frac{296+169\sqrt{6}}{1800}$ & $\frac{88+7\sqrt{6}}{360}$ & $\frac{-2-3\sqrt{6}}{225}$ \\
Internal stage & 3 & $1$ & $\frac{16-\sqrt{6}}{36}$ & $\frac{16+\sqrt{6}}{36}$ & $\frac{1}{9}$ \\
Final update & -- & -- & $b_1=\frac{16-\sqrt{6}}{36}$ & $b_2=\frac{16+\sqrt{6}}{36}$ & $b_3=\frac{1}{9}$
\end{tabular}
\end{ruledtabular}
\end{table*}

For DAE, with stage states $Y_i$ and stage derivatives $K_i \approx \dot{y} (t
+ c_i h)$, the collocation equations are
\begin{equation}
  M K_i = f (t_n + c_i h, Y_i), i = 1, 2, 3.
\end{equation}
We work with stage displacements $Z_i : = h \sum_j a_{i j} K_j$ and stack them
as $Z \in \mathbb{R}^{3 \times n}$. A simplified Newton iteration (Jacobian $J
= \partial f / \partial y$ frozen per step, and numerical differentiation
$\frac{d f}{d x} = \lim_{h \rightarrow 0}  \frac{f (x + h) - f (x)}{h}$ is use
to calculate the jacobian matrix {\cite{1974est}}) would yield a $3 n \times 3
n$ block system
\begin{equation}
  (I_3 \otimes M - h A \otimes J) \Delta K = -\mathcal{R},
\end{equation}
where $I_3$ is the $3 \times 3$ identity matrix, $A = [a_{11}, \cdots ; \cdots ;
\cdots, a_{33}]$,
\begin{equation}
  \mathcal{R}= \left(\begin{array}{c}
    M K_1 - f (t_1 + c_1 h, Y_1)\\
    M K_2 - f (t_2 + c_2 h, Y_2)\\
    M K_3 - f (t_2 + c_3 h, Y_3)
  \end{array}\right) .
\end{equation}
The implementation diagonalizes the $3\times 3$ stage-coupling matrix once,
\begin{equation}
  A = T \Lambda T^{- 1},
\end{equation}
where $\Lambda = \tmop{diag} (\lambda_{\tmop{real}}, \lambda, \bar{\lambda})$
with one real eigenvalue and a complex-conjugate pair, $T$ is fixed change of
basis:
\begin{eqnarray}
  \lambda_{\tmop{real}} & = & \frac{1}{3 + 3^{2 / 3} - 3^{1 / 3}} : =
  \frac{1}{\mu_{\tmop{real}}}, \nonumber\\
  \lambda, \bar{\lambda} & = & \frac{1}{3 + \frac{1}{2} (3^{1 / 3} - 3^{2 /
  3}) \mp \frac{i}{2} (3^{5 / 6} + 3^{7 / 6})} \assign
  \frac{1}{\mu_{\lambda}}, \frac{1}{\mu_{\bar{\lambda}}}, \nonumber\\
  \Lambda & = & \left(\begin{array}{ccc}
    \lambda_{\tmop{real}} & 0 & 0\\
    0 & \lambda & 0\\
    0 & 0 & \bar{\lambda}
  \end{array}\right), \\
  T & = & \left(\begin{array}{ccc}
    \frac{3}{32} - \frac{\sqrt{6}}{48} & - \frac{3}{32} - \frac{\sqrt{6}}{48}
    & \frac{1}{16}\\
    \frac{1}{4} & \frac{1}{4} & - \frac{3}{8}\\
    1 & 1 & 0
  \end{array}\right), \nonumber\\
  T^{- 1} & = & \left(\begin{array}{ccc}
    \frac{16}{3} & - \frac{4}{3} & \frac{8}{3}\\
    - \frac{16}{3} & \frac{4}{3} & \frac{4}{3}\\
    \frac{8}{3} & - \frac{20}{3} & \frac{8}{3}
  \end{array}\right) . \nonumber
\end{eqnarray}
Changes stage variables $W = T^{- 1} Z$ (so $Z = T W$), $W = W_0 + (W_1 + i
W_2)$, in these coordinates, the Newton system decouples into three $n \times
n$ systems:
\begin{eqnarray}
  h_s \left( \frac{\mu_{\ell}}{h} M - J \right) \Delta W_{\ell} & = &
  r_{\ell}, 
\end{eqnarray}
where
\begin{align}
  \mu_{\ell} &= 1 / \lambda_{\ell}, \qquad \ell \in \{ \tmop{real}, \lambda, \bar{\lambda} \}, \\
  r_{\tmop{real}} &= S \left( F^{\top} T_{\tmop{real}} - \frac{\mu_{\tmop{real}}}{h} M W_0 \right), \\
  r_{\tmop{complex}} &= S \left( F^{\top} T_{\tmop{complex}} - \frac{\mu_{\tmop{complex}}}{h} M (W_1 + i W_2) \right), \\
  F &= [f (t_1 + c_1 h, Y_1) ; f (t_2 + c_2 h, Y_2) ; \notag\\
    &\qquad f (t_3 + c_3 h, Y_3)].
\end{align}
and $h_s, S$ is diagonal scaling:
\begin{equation}
  S = \tmop{diag} (h_s) .
\end{equation}
Because two eigenvalues form a complex-conjugate pair, only two LU
factorizations are needed per step: one real and one complex; the complex
update is split into real/imaginary parts and mapped back by $T$.

After Newton converges, an embedded order-3 error is computed by reusing the
real LU:

\begin{equation}
  \left( \frac{\mu_{\text{real}}}{h} \text{ } M - J \right) \text{ } e = f
  (t_n, y_n) + \frac{1}{h} M \text{ } Z_E,
\end{equation}

with a fixed linear combination $Z_E$ of stages. The scaled norm $\parallel e
\parallel_{\text{sc}}$ is compared to 1 (optional handling of algebraic
components: scale/zero in the norm). If $\parallel e \parallel_{\text{sc}}
\leq 1$ the step is accepted, else rejected.

There is one step size $h$ per step. A {\tmem{Gustafsson}}-style predictor
controls it:
\begin{equation}
  h_{n + 1} = h_n \times \left( \frac{\tmop{tol}}{\tmop{error}} \right)^{1 /
  (p + 1)},
\end{equation}
On acceptance, we deal next step $h_{n + 1}$ with
\begin{equation}
  h_{n + 1} = h_n \times \varrho_f
\end{equation}
where
\[ \varrho_f = \min \{ \varrho_{\max}, \varpi \cdummy \varrho \}, \]
on rejection, we use
\begin{equation}
  \varrho_f = \max \{ \varrho_{\min}, \varpi \cdummy \varrho \}
\end{equation}
to renew recent time step $h_n = h_n \times \varrho_f$, where $\varrho_f$ is
the time updating factor, calculated by factor $\varrho$ and safety factor
$\varpi$.

$\varrho$ is defined by a prediction multiplier
\begin{equation}
  \varrho = \min \left\{ 1, \frac{h_n}{h_{n - 1}} \times \left(
  \frac{\tmop{error}_{n - 1}}{\tmop{error}_n} \right)^{1 / 4} \right\} \cdot
  \tmop{error}^{- 1 / 4},
\end{equation}
and $\varpi$ defined w.r.t. newton iteration step:
\begin{equation}
  \varpi = 0.9 \times \frac{2 \times k_{\max} + 1}{2 \times k_{\max} + k},
\end{equation}
where $k_{\max}$ is the maximum newton step we expect.

The adaptive Radau IIA stepping procedure used in our implementation is summarized in Algorithm~1.

\begin{algorithm}[t]
\caption{Adaptive Radau IIA step with frozen Jacobian}
\DontPrintSemicolon
\SetAlgoLined
\KwIn{Current state $(t,y)$, step size $h$, mass matrix $M$, frozen Jacobian $J$}
\KwData{$\tmop{Tol}_{\tmop{newton}}=10^{-3}$, $k_{\max}=6$, $\varrho_{\max}=10$, $\varrho_{\min}=0.2$}
\Repeat{step accepted}{
  \If{$h$ is too small}{fail}
  Compute $t_i=t+c_i h$ and initialize $Z_0=0$ (or extrapolate from dense output)\;
  Build $H_{\tmop{real}}=h_{\tmop{scaled}}\bigl((\mu_{\tmop{real}}/h)M-J\bigr)$ and $H_{\tmop{cplx}}=h_{\tmop{scaled}}\bigl((\mu_{\tmop{cplx}}/h)M-J\bigr)$\;
  Factorize $\tmop{LU}_{\tmop{real}}=\tmop{LU}(H_{\tmop{real}})$ and $\tmop{LU}_{\tmop{cplx}}=\tmop{LU}(H_{\tmop{cplx}})$\;
  Set $W=T^{-1}Z_0$\;
  \For{$k=1,\ldots,k_{\max}$}{
    Evaluate $F_i=f(t_i,y+Z_i)$\;
    Form $r_{\tmop{real}}=S\bigl(F^{\top}T^{-1}_{\tmop{real}}-(\mu_{\tmop{real}}/h)MW_0\bigr)$\;
    Form $r_{\tmop{cplx}}=S\bigl(F^{\top}T^{-1}_{\tmop{cplx}}-(\mu_{\tmop{cplx}}/h)M(W_1+iW_2)\bigr)$\;
    Solve $\Delta W_0=\tmop{solve}(\tmop{LU}_{\tmop{real}},r_{\tmop{real}})$\;
    Solve $\Delta (W_1+iW_2)=\tmop{solve}(\tmop{LU}_{\tmop{cplx}},r_{\tmop{cplx}})$\;
    Update $W \leftarrow W+\Delta W$ and $Z \leftarrow TW$\;
    \If{$\|\Delta W\|_{\tmop{scaled}}<\tmop{Tol}_{\tmop{newton}}$}{break}
  }
  \If{Newton iteration failed}{
    $h \leftarrow \max\{\varrho_{\min},\varpi\cdot\varrho\}h$\;
    continue
  }
  Compute the embedded error estimate $e=\tmop{solve}(\tmop{LU}_{\tmop{real}},f(t,y)+(1/h)MZ_E)$\;
  \If{$\|e\|_{\tmop{sc}}\le 1$}{
    Accept the step: $y\leftarrow y+h\sum b_i K_i(Z)$ and $t\leftarrow t+h$\;
    Set $h_{\tmop{next}}\leftarrow \min\{\varrho_{\max},\varpi\cdot\varrho\}h$ and return\;
  }
  $h \leftarrow \max\{\varrho_{\min},\varpi\cdot\varrho\}h$\;
}
\end{algorithm}

\section{Results and Verification}

\subsection{Spatial convergence}

Due to non-uniform mesh established, we use the representative mesh size $h =
\frac{L}{N}$ (except ghost points) to measure spatial step. Because the exact
solution is not available, a reference solution is computed on a very fine
grid with $N^{\tmop{ref}} = 3200$, and $N \in \{ 50, 100, 200, 400, 800, 1600
\}$ for each coarser grid. On the non-uniform grid we define the discrete
$L^2$--error by the trapezoidal rule
\begin{equation}
  E_u (h) = \left( \sum_{i = 1}^N \omega_i  | u_h (x_i) - u^{\tmop{ref}} (x_i)
  |^2 \right)^{1 / 2},
\end{equation}
where $\omega_i$ are the standard trapezoidal weights (control volume) built
from the local mesh widths, and only interior nodes are included in the sum.

\begin{figure}[t]
\includegraphics[width=\columnwidth]{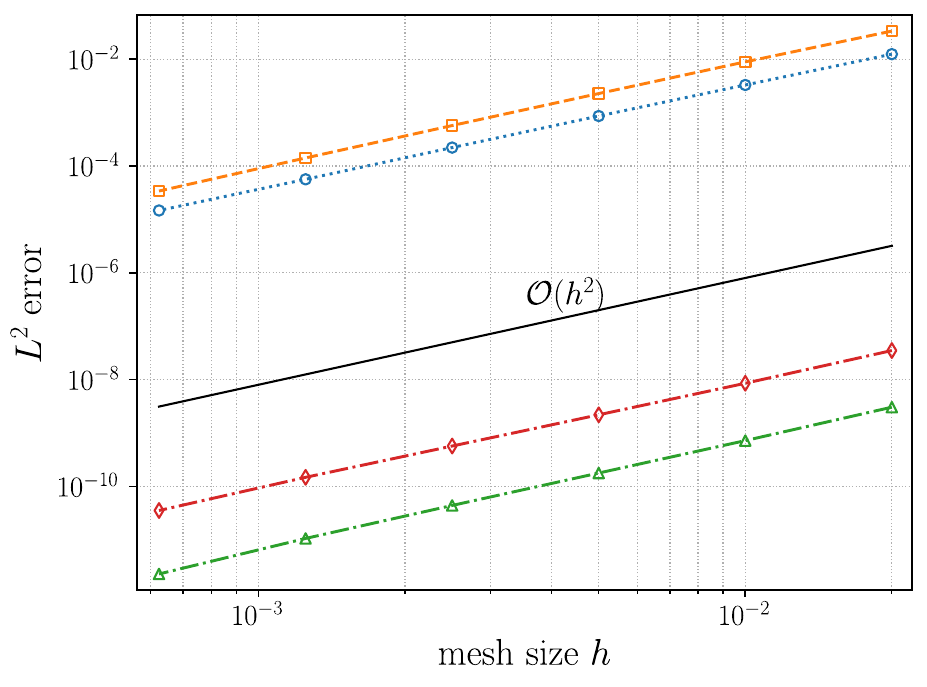}
\caption{$L^2$ error versus mesh size $h$ for $P$ (blue), $\phi$ (orange), $n$ (green), and $p$ (red), with $\mathcal{O}(h^2)$ reference line (black).}
\label{L2h}
\end{figure}

Figure \ref{L2h} shows $L^2$ error $E_u (h)$ versus $h$ in log--log scale for
$u = P, \phi, n, p$ yields straight lines with slopes close to two, this
demonstrates that the proposed spatial discretization is globally
{\tmem{second-order accurate}} for all four variables.

\subsection{Temporal convergence}

\begin{figure}[t]
\includegraphics[width=\columnwidth]{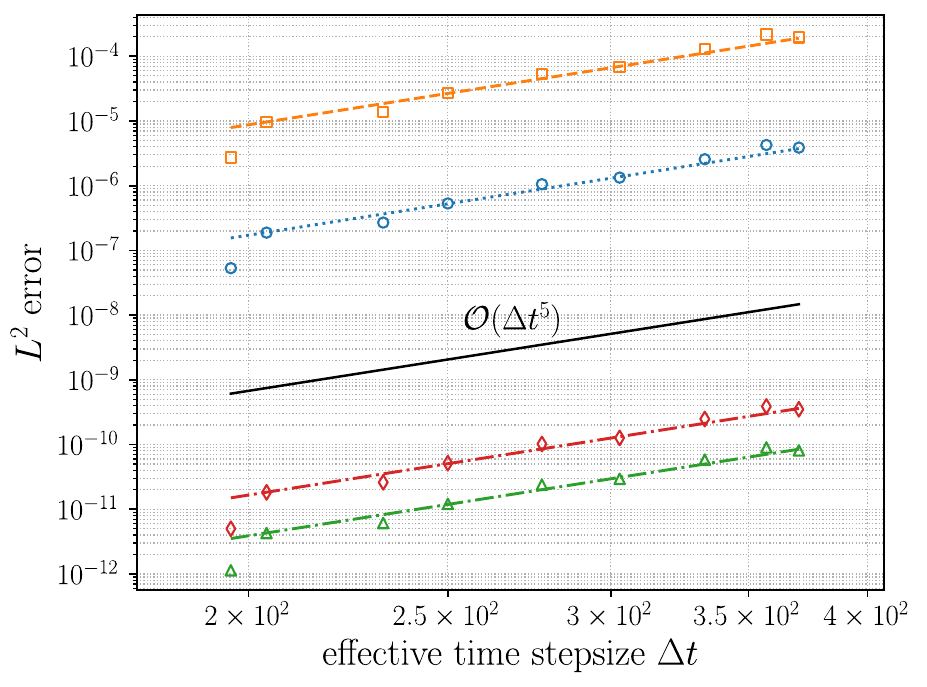}
\caption{$L^2$ error versus effective time stepsize $\Delta t$ for $P$ (blue), $\phi$ (orange), $n$ (green), and $p$ (red), with $\mathcal{O}(t^5)$ reference line (black).}
\label{L2t}
\end{figure}

The steps are non-uniform because adaptive time stepping controlled by
relative and absolute tolerances {\tmem{rtol}} and {\tmem{atol}} via
{\tmem{Gustafsson}}-style predictor. Thus, we characterize the temporal
resolution by an {\tmem{effective time-step size}}. Let $\{ t_n \}_{n = 0}^M$
be the internal time grid produced by the solver during the second stage, and
define
\begin{equation}
  \Delta t^{\tmop{eff}} = \frac{t_M - t_0}{M},
\end{equation}
where $M$ is the number of internal steps. We use a very tight tolerance pair
({\tmem{rtol}} $= 10^{- 6}$ and {\tmem{atol}} $= 10^{- 9}$) as reference
solution and a sequence of runs are performed for decreasing tolerances
{\tmem{rtol}} $\in \{ 10^{- 2}, 5 \times 10^{- 3}, 2.5 \times 10^{- 3},
\cdots, 3.90625 \times 10^{- 5} \}$ with {\tmem{atol}} $= 10^{- 3}
\times${\tmem{rtol}}, which computed on the same spatial grid ($N = 200$) and
results in a corresponding sequence of decreasing effective time stepsizes
$\Delta t^{\tmop{eff}}$. For each run with a given $\Delta t_{\tmop{eff}}$ we
then measure, at the final time $T$, the temporal error in the spatial $L^2$,
\begin{equation}
  E_u (\Delta t^{\tmop{eff}}) = \left( \sum_{i = 1}^N \omega_i  | u_h (x_i, T)
  - u^{\tmop{ref}} (x_i, T) |^2 \right)^{1 / 2},
\end{equation}
where the quadrature weights $\omega_i$ are the same trapezoidal weights as in
the spatial convergence study. Figure \ref{L2t} plots $L^2$ error $E_u (\Delta
t_{\tmop{eff}})$ versus $\Delta t_{\tmop{eff}}$ in log--log scale together
with a reference line of slope five, corresponding to $O (\Delta t^5)$.

The numerical results show that, as the effective time stepsize $\Delta t$
decreases, the $L^2$ error for all components $P$, $\phi$, $n$, and $p$ decays
along straight lines that are essentially parallel to the reference curve $O
(\Delta t^5)$ over a wide intermediate range of $\Delta t$, which confirms the
expected {\tmem{fifth--order temporal accuracy}} of the Radau scheme for this
problem. For the largest stepsizes the curves deviate slightly from this
slope, indicating pre--asymptotic behavior before the leading $\Delta t^5$
term dominates; for sufficiently small stepsizes one would similarly expect a
departure from the ideal $\Delta t^5$ scaling once the temporal error becomes
comparable to spatial and algebraic errors, although this regime is not yet
pronounced in the present simulations.

\subsection{Computational Efficiency}

We now compare the computational efficiency of our numerical method, IonMonger 2.0,
Driftfusion, OghmaNano, and COMSOL using the same benchmark. The benchmark is a
standard one-dimensional, three-layer electronic Poisson--drift--diffusion
structure: a 400~nm absorber sandwiched between a 100~nm electron-transport
layer and a 200~nm hole-transport layer. Mobile ions, traps, and excitons were
disabled. The device parameters are listed in Table~\ref{tab:timing-case}. The
voltage was scanned from 0 to 1.2~V at 0.1~V~s$^{-1}$. Mesh targets of 80, 160,
$\ldots$, 800 total nodes were used. At each target, we performed one warm-up
run followed by ten timed runs for the implementations with ten repeats
available; IonMonger 2.0 was timed over twenty runs because its timing file
contains twenty repeats. Setup and mesh construction were excluded. Reported
values are median times for the complete forward $J$--$V$ scan; error bars
indicate the first and third quartiles.

\begin{table*}[t]
\caption{Device parameters for the common electronic benchmark used for the timing comparison.}
\label{tab:timing-case}
\begin{ruledtabular}
\begin{tabular}{ccllll}
Quantity & Symbol & ETL & Absorber & HTL & Unit \\
\colrule
Layer thickness & $d$ & 100 & 400 & 200 & nm \\
Relative permittivity & $\varepsilon_r$ & 10.0 & 24.1 & 3.0 & -- \\
Electron mobility & $\mu_n$ & 3.89 & 66.2 & 0.389 & $\mathrm{cm^2\,V^{-1}\,s^{-1}}$ \\
Hole mobility & $\mu_p$ & 3.89 & 66.2 & 0.389 & $\mathrm{cm^2\,V^{-1}\,s^{-1}}$ \\
Electron lifetime & $\tau_n$ & -- & $3\times10^{-9}$ & -- & s \\
Hole lifetime & $\tau_p$ & -- & $3\times10^{-7}$ & -- & s \\
Surface recombination velocity & $S$ & $10^7$ & -- & $10^7$ & $\mathrm{cm\,s^{-1}}$ \\
\end{tabular}
\end{ruledtabular}
\end{table*}

The optical generation in the absorber follows a Beer--Lambert profile. With
$x_0=10^{-5}$~cm and $x_1=5\times10^{-5}$~cm denoting the left and right
boundaries of the absorber, respectively,
\begin{equation}
G(x)=\begin{cases}
\Phi\alpha\exp[-\alpha(x-x_0)], & x_0\le x\le x_1,\\
0, & \text{otherwise},
\end{cases}
\label{eq:beer-lambert}
\end{equation}
where $\Phi=1.4\times10^{17}$~cm$^{-2}$s$^{-1}$ is the incident photon flux
and $\alpha=1.3\times10^5$~cm$^{-1}$ is the absorption coefficient.

The differences in computational efficiency arise from the numerical
implementation used in each framework: this work uses damped Newton iterations
with sparse direct linear solves; IonMonger 2.0 uses its finite-element
differential-algebraic formulation and \texttt{ode15s}; Driftfusion uses MATLAB
\texttt{pdepe}; OghmaNano uses its own device solver; and COMSOL uses its
commercial finite-element solver.

\begin{figure}[t]
\centering
\includegraphics[width=0.95\columnwidth]{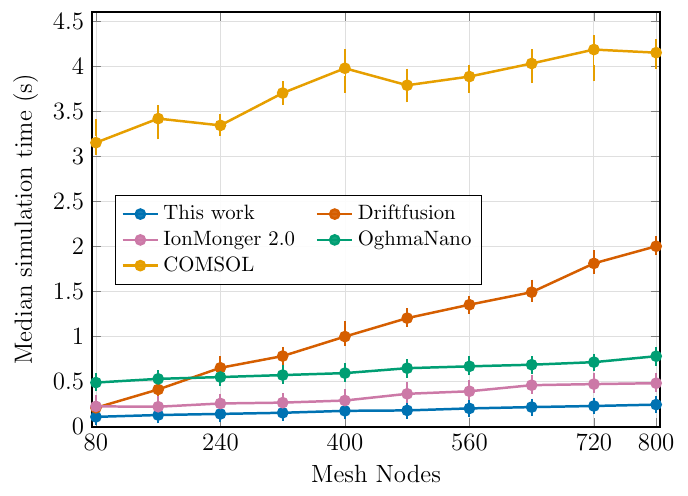}
\caption{Median simulation time versus the number of mesh nodes for five
solver implementations on the common electronic benchmark. Markers indicate
the median of the timed runs, and asymmetric error bars represent the first and
third quartiles.}
\label{timing-plot}
\end{figure}

As shown in Figure~\ref{timing-plot}, our method has the lowest median runtime at
every mesh size tested in this benchmark. The median time across the ten mesh
sizes is 0.178~s for this work, compared with 0.343~s for IonMonger 2.0,
0.624~s for OghmaNano, 1.092~s for Driftfusion, and 3.763~s for COMSOL. Thus,
the other implementations take approximately 1.93, 3.50, 6.14, and 21.15 times
as long as our method, respectively. These results show that our implementation
is the fastest among the tested implementations under the stated benchmark.
However, they should not be interpreted as a universal ranking, because runtime
depends on the model, mesh, tolerances, hardware, and solver settings. Such
differences are part of the practical comparison; therefore, the results are
reported as a benchmark of complete program workflows rather than as a
comparison of a single isolated linear-algebra kernel.

\subsection{Numerical Stability}

We now assess the solver's numerical stability through a parameter-convergence
study, in which the solver is tested with different numbers of mesh nodes while
varying the applied voltage, temperature, and contact doping. The meshes contain
100, 200, 300, 400, 500, 600, or 700 nodes; each reported solution is compared
against a separate 1401-node reference solution. The voltage sweep uses
$V_{\mathrm{app}}=-1$, $0$, and $+1$~V at 300~K, with $10^{18}$~cm$^{-3}$ regular
contact doping and one-sun Beer--Lambert generation profile. The temperature
sweep uses 100, 300, and 500~K at zero bias under dark equilibrium, and the
doping sweep uses contact doping of $10^{18}$, $10^{20}$, $10^{21}$, and
$10^{22}$~cm$^{-3}$ at 300~K, zero bias, and one-sun generation. The
$10^{22}$~cm$^{-3}$ case is deliberately an extreme numerical stress test, not a
claim about a typical device composition. For each mesh, we interpolate the
solution onto the reference grid and compute normalized relative $L^2$ errors for
the electrostatic potential, electron density, and hole density. We summarize
these errors by
\begin{equation}
 \mathrm{error}_{\mathrm{sum}}=\left[\mathrm{error}_{\phi}^{2}+\mathrm{error}_{n}^{2}+\mathrm{error}_{p}^{2}\right]^{1/2}.
\end{equation}

\begin{figure*}[t]
\centering
\includegraphics[width=0.98\textwidth]{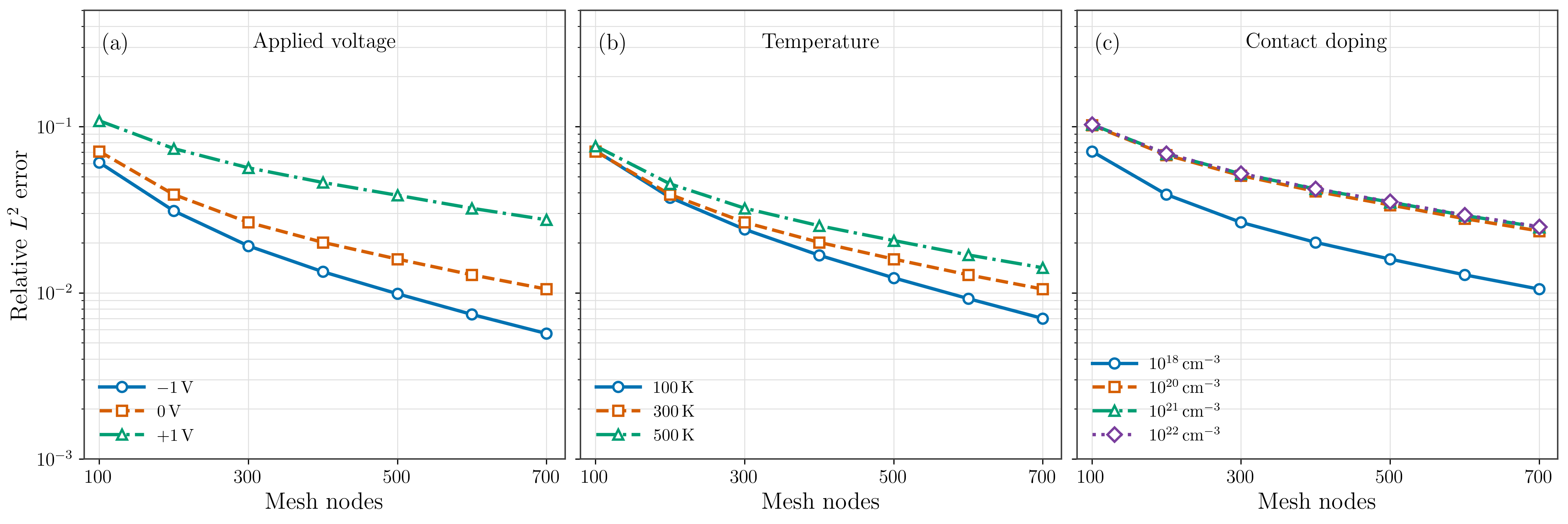}
\caption{Relative $L^2$ error versus the number of mesh nodes for the voltage,
temperature, and contact-doping sweeps. The errors
are computed relative to the corresponding 1401-node interface-fitted reference
solution.}
\label{stability-convergence}
\end{figure*}

All 70 parameter--mesh combinations and all ten 1401-node reference calculations
converged, with finite and positive carrier densities. Figure~\ref{stability-convergence}
shows that $\mathrm{error}_{\mathrm{sum}}$ decreases systematically as the mesh is refined,
i.e., as the number of nodes increases. At 700 nodes, the largest $\mathrm{error}_{\mathrm{sum}}$
values are $2.75\times10^{-2}$, $1.41\times10^{-2}$, and $2.50\times10^{-2}$
for the voltage, temperature, and doping sweeps, respectively
(Table~\ref{tab:stability-convergence}). The largest reported infinity-norm
nonlinear residual at 700 nodes is $1.57\times10^{-10}$, and all reported runs
satisfied the convergence criterion. These results demonstrate mesh consistency
and robustness over the tested parameter ranges. They also clarify the scope of
this test: it verifies the robustness of the steady-state nonlinear solution and
spatial convergence, whereas the high-order Radau IIA time integrator is assessed
in the temporal-convergence study described above.

\begin{table*}[t]
\caption{Parameter-convergence study used to assess numerical robustness. Each
mesh solution is compared with a corresponding 1401-node reference; the reported
error is the maximum $\mathrm{error}_{\mathrm{sum}}$ at 700 nodes within the indicated sweep.}
\label{tab:stability-convergence}
\begin{ruledtabular}
\small
\begin{tabular}{lllll}
Sweep & Values & Fixed conditions & Illumination & $\max \mathrm{error}_{\mathrm{sum}}$ at 700 nodes \\
\colrule
Applied voltage & $-1,0,+1$~V & $T=300$~K, $N_C=10^{18}$~cm$^{-3}$ & 1 sun & $2.75\times10^{-2}$ \\
Temperature & $100,300,500$~K & $V_{\mathrm{app}}=0$, $N_C=10^{18}$~cm$^{-3}$ & dark equilibrium & $1.41\times10^{-2}$ \\
Contact doping & $10^{18}$--$10^{22}$~cm$^{-3}$ & $T=300$~K, $V_{\mathrm{app}}=0$ & 1 sun & $2.50\times10^{-2}$ \\
\end{tabular}
\end{ruledtabular}
\end{table*}

\subsection{\texorpdfstring{Verification on a $p$-$n$ junction}{Verification on a p-n junction}}

As a first steady--state test we consider an abrupt, symmetric $p$-$n$
junction of total length $L = 400 \text{ } \tmop{nm}$. The structure consists
of an $n$-type region of thickness $L_n = 200 \text{ } \tmop{nm}$ followed by
a $p$-type region of the same thickness. The donor and acceptor densities are
$N_D = N_A = 2.9 \times 10^{22} \text{ } m^{- 3}$, the intrinsic carrier
concentration is $n_i = 3.5 \times 10^{16} \text{ } m^{- 3}$, and the
temperature is $T = 298 \text{ } K$, so that the thermal voltage is $V_t = k_B
T / q \approx 2.57 \times 10^{- 2} \text{ } V$. The geometric interface at
which the doping changes from $N_D$ to $N_A$ is referred to as the
metallurgical junction and is located at $x_j = L_n$.

For an abrupt $p$-$n$ junction, the classical depletion approximation predicts
the built--in voltage

\begin{equation}
  V_{\tmop{bi}} = V_t \ln \nospace \text{ {\nocomma}} \left( \frac{N_D
  N_A}{n_i^2} \right),
\end{equation}

and the depletion widths in the $n$- and $p$-regions

\begin{align}
  W_n &= \sqrt{\frac{2 \varepsilon_s}{q} \text{ } \frac{V_{\tmop{bi}} N_A}{N_D (N_D + N_A)}}, \notag\\
  W_p &= \sqrt{\frac{2 \varepsilon_s}{q} \text{ } \frac{V_{\tmop{bi}} N_D}{N_A (N_D + N_A)}}.
\end{align}

where $\varepsilon_s$ is the semiconductor permittivity. With the above
parameters we obtain $V_{\tmop{bi}} \approx 0.700 \text{ } V$ and $W_n \approx
W_p \approx 73.0 \text{ } \tmop{nm}$, so that the total theoretical depletion
width is $W_{\tmop{th}} = W_n + W_p \approx 146.0 \text{ } \tmop{nm}$.

Let $x^{\prime} = x - x_j$ be the coordinate with origin at the metallurgical
junction. Within the depletion region $- W_n \leq x^{\prime} \leq W_p$, the
depletion approximation assumes that only ionized dopants contribute to the
space charge. The electrostatic potential (referenced to its value $\phi_L$ in
the left neutral region) is then

\begin{equation}
\begin{split}
\phi_{\tmop{dep}} (x) - \phi_L
&= - \dfrac{q N_D}{2 \varepsilon_s} (x^{\prime} + W_n)^2,
\qquad - W_n \le x^{\prime} \le 0,\\[4pt]
\phi_{\tmop{dep}} (x) - \phi_L
&= - \dfrac{q N_D W_n^2}{2 \varepsilon_s} +{},\\
&\qquad \dfrac{q N_A}{2 \varepsilon_s} \bigl[(x^{\prime} - W_p)^2 - W_p^2\bigr],\\
&\qquad 0 \le x^{\prime} \le W_p.
\end{split}
\end{equation}

and the electric field is piecewise linear,\vspace{-0.25\baselineskip}
\begin{equation}
\begin{aligned}
E_{\tmop{dep}} (x) &= - \frac{d \phi_{\tmop{dep}}}{d x} \\
&= \left\{\renewcommand{\arraystretch}{1.15}
\begin{array}{@{}l@{\qquad\qquad}l@{}}
\dfrac{q N_D}{\varepsilon_s} (x^{\prime} + W_n), & - W_n \le x^{\prime} \le 0,\\[7pt]
\dfrac{q N_A}{\varepsilon_s} (W_p - x^{\prime}), & 0 \le x^{\prime} \le W_p,\\[7pt]
0, & |x^{\prime}| > W_{n,p}.
\end{array}
\right.
\end{aligned}
\end{equation}

From the numerical drift--diffusion solution we reconstruct the dimensional
carrier profiles $n (x), p (x)$ and potential $\phi (x)$. To define a
depletion width from the numerical data, we use a majority--carrier threshold
criterion. Denoting by $x_j$ the metallurgical junction, we set
\begin{eqnarray}
  r_n (x) = \frac{n (x)}{N_D (x)}, &  & x < x_j ; \\
  r_p (x) = \frac{p (x)}{N_A (x)}, &  & x > x_j, 
\end{eqnarray}
and, for a given threshold $\eta \in (0, 1)$,

\begin{align}
  x_n^{\tmop{num}} &= \min \{ x < x_j : \text{ } r_n(x) \le \eta \}, \notag\\
  x_p^{\tmop{num}} &= \max \{ x > x_j : \text{ } r_p(x) \le \eta \}.
\end{align}

with numerical depletion width

\begin{equation}
  W_{\tmop{dep}}^{\tmop{num}} = x_p^{\tmop{num}} - x_n^{\tmop{num}} .
\end{equation}

\begin{figure}[t]
\includegraphics[width=\columnwidth]{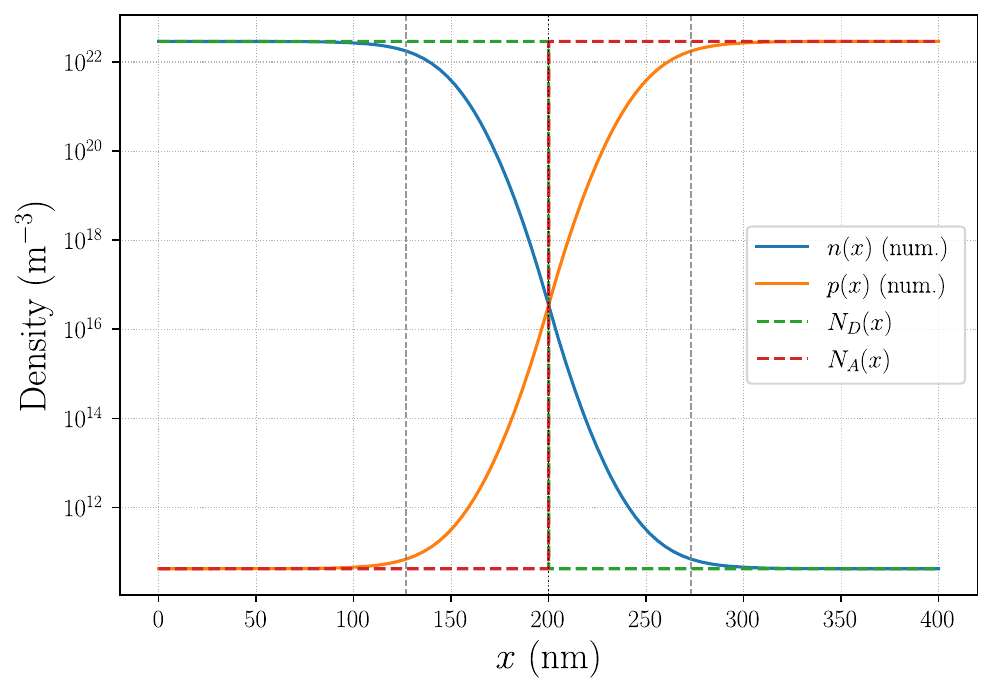}
\caption{Equilibrium carrier and doping profiles for the abrupt $p$-$n$ junction. Solid lines show the numerical electron and hole densities $n(x)$ and $p(x)$; dashed lines indicate the ionized donor and acceptor concentrations $N_D(x)$ and $N_A(x)$. The vertical dotted line marks the metallurgical junction, and the grey dashed lines indicate the depletion edges obtained from the majority--carrier criterion with $\eta = 0.67$.}
\label{PNc}
\end{figure}
In this work we choose $\eta = 0.67$. We have verified that the extracted
width varies only weakly when $\eta$ is varied in the range $[0.5, 0.8]$, so
that $W_{\tmop{dep}}^{\tmop{num}}$ is not sensitive to the precise choice of
the threshold.

Using this definition we obtain $W_{\tmop{dep}}^{\tmop{num}} \approx 146.4
\text{ } \tmop{nm}$, with depletion edges at $x \approx 126.8 \text{ }
\tmop{nm}$ and $x \approx 273.2 \text{ } \tmop{nm}$. These values are in
excellent agreement with the analytical predictions $x_n^{\tmop{th}} = x_j -
W_n \approx 127.0 \text{ } \tmop{nm}$ and $x_p^{\tmop{th}} = x_j + W_p \approx
273.0 \text{ } \tmop{nm}$. The built--in voltage extracted from the numerical
potential, $\phi (L) - \phi (0) \approx - 0.700 \text{ } V$, also matches the
Shockley formula to within numerical precision.

\begin{figure}[t]
\includegraphics[width=\columnwidth]{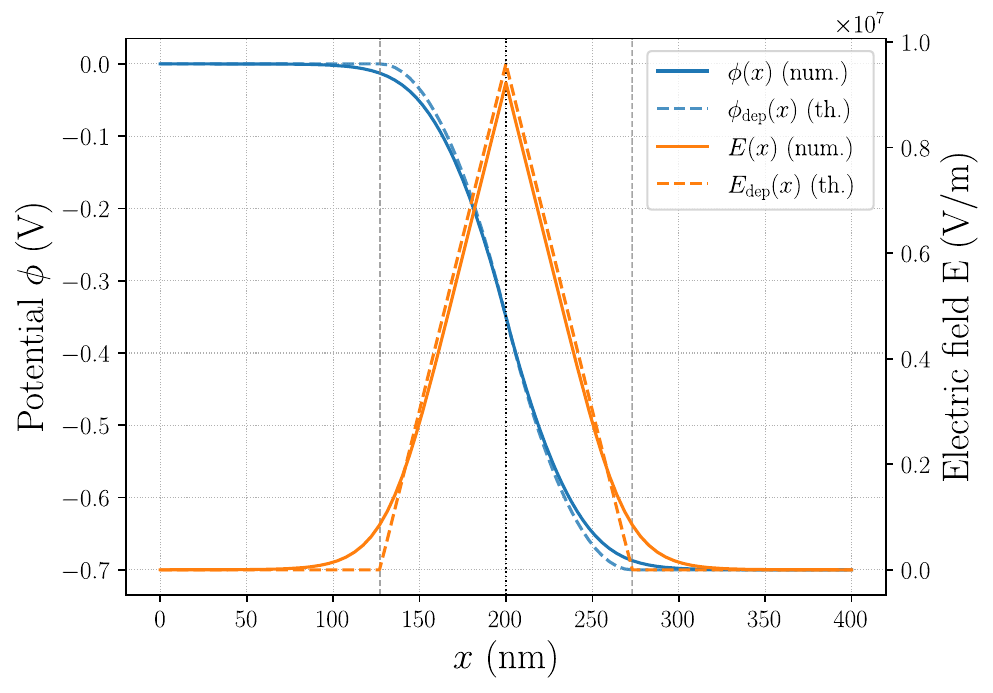}
\caption{Electrostatic potential and electric field in the $p$-$n$ junction. Solid lines show the numerical drift--diffusion solution $\phi(x)$ and $E(x)$; dashed lines show the analytical depletion--approximation profiles $\phi_{\tmop{dep}}(x)$ and $E_{\tmop{dep}}(x)$. The metallurgical junction and depletion boundaries are indicated as in Figure~\ref{PNc}. The close agreement demonstrates that the numerical scheme reproduces the classical equilibrium $p$-$n$ junction solution.}
\label{PNv}
\end{figure}
Figure \ref{PNc} shows the equilibrium electron and hole profiles together
with the piecewise--constant doping densities. The numerically defined
depletion region based on the majority--carrier criterion almost coincides
with the analytical depletion region. Figure \ref{PNv} compares the numerical
potential and electric field with the depletion--approximation solutions
$\phi_{\tmop{dep}} (x)$ and $E_{\tmop{dep}} (x)$. The numerical scheme
reproduces the textbook $p$-$n$ junction behavior: a nearly flat potential in
the neutral regions, a parabolic potential drop and triangular electric field
in the depletion region, and a peak field at the metallurgical junction.

\subsection{Validation against existing software}

We compare the present drift--diffusion solver with Driftfusion, IonMonger 2.0,
COMSOL, and OghmaNano using the same PNDIT-F3N/D18:L8-BO/PEDOT:PSS device,
material parameters (Table~\ref{table-compare}), and common voltage range
$0$--$0.9$~V. The five methods therefore evaluate the same device-level problem;
the comparison is not a ranking of unrelated solver configurations. From each
computed $J$--$V$ curve we extract $V_{\mathrm{oc}}$, $J_{\mathrm{sc}}$,
$(V_{\mathrm{mpp}},J_{\mathrm{mpp}})$, $P_{\max}$, FF, and PCE
(Table~\ref{table-metrics}).

\begin{table*}[t]
\caption{Key model parameters for the PNDIT-F3N/D18:L8-BO/PEDOT:PSS organic solar cell used in the five-method comparison and in this work.}
\label{table-compare}
\begin{ruledtabular}
\small
\begin{tabular}{llcccc}
Quantity & Symbol & PNDIT-F3N & D18:L8-BO & PEDOT:PSS & Unit \\
\colrule
Layer thickness & $d$ & 5 & 120 & 20 & nm \\
Relative permittivity & $\varepsilon_r$ & 3.0 & 3.0 & 3.0 & -- \\
Electron diffusion coefficient & $D_n$ & $1.0\times10^{-3}$ & $3.9\times10^{-9}$ & $1.0\times10^{-3}$ & $\mathrm{m^2\,s^{-1}}$ \\
Hole diffusion coefficient & $D_p$ & $1.0\times10^{-3}$ & $3.9\times10^{-9}$ & $1.0\times10^{-3}$ & $\mathrm{m^2\,s^{-1}}$ \\
Conduction-band DOS & $N_c$ & $1.0\times10^{26}$ & $1.0\times10^{26}$ & $1.0\times10^{26}$ & $\mathrm{m^{-3}}$ \\
Valence-band DOS & $N_v$ & $1.0\times10^{26}$ & $1.0\times10^{26}$ & $1.0\times10^{26}$ & $\mathrm{m^{-3}}$ \\
Electron affinity & $\Phi_{\tmop{EA}}$ & $-4.2$ & $-4.2$ & $-4.2$ & eV \\
Ionisation potential & $\Phi_{\tmop{IP}}$ & $-5.48$ & $-5.48$ & $-5.48$ & eV \\
Equilibrium Fermi level & $E_{F0}$ & $-4.245$ & $-4.84$ & $-5.435$ & eV \\
Bulk bimolecular recomb. & $\beta$ & 0 & $1.7\times10^{-17}$ & 0 & $\mathrm{m^3\,s^{-1}}$ \\
Uniform generation rate & $G$ & 0 & $1.358\times10^{28}$ & 0 & $\mathrm{m^{-3}\,s^{-1}}$ \\
Left boundary electron recomb. velocity & $v_{n,l}$ & $1.0\times10^{5}$ & -- & -- & $\mathrm{m^2\,s^{-1}}$ \\
Left boundary hole recomb. velocity & $v_{p,l}$ & $1.0\times10^{5}$ & -- & -- & $\mathrm{m^2\,s^{-1}}$ \\
Right boundary electron recomb. velocity & $v_{n,r}$ & -- & -- & $1.0\times10^{5}$ & $\mathrm{m^2\,s^{-1}}$ \\
Right boundary hole recomb. velocity & $v_{p,r}$ & -- & -- & $1.0\times10^{5}$ & $\mathrm{m^2\,s^{-1}}$
\end{tabular}
\end{ruledtabular}
\end{table*}

\begin{table*}[t]
\caption{Extracted photovoltaic metrics from the $J$--$V$ curves for the same
device and common voltage protocol: this work, Driftfusion, IonMonger 2.0,
COMSOL, and OghmaNano.}
\label{table-metrics}
\begin{ruledtabular}
\small
\begin{tabular}{lccccccc}
 & $V_{\mathrm{oc}}$ (V) & $J_{\mathrm{sc}}$ ($\mathrm{mA}/\mathrm{cm}^2$) & $V_{\mathrm{mpp}}$ (V) & $J_{\mathrm{mpp}}$ ($\mathrm{mA}/\mathrm{cm}^2$) & $P_{\max}$ ($\mathrm{mW}/\mathrm{cm}^2$) & FF & PCE (\%) \\
\colrule
This work & 0.84876884 & 25.36517 & 0.7245000 & 23.07638095 & 16.71883800 & 0.776566775 & 16.71883800 \\
Driftfusion & 0.84866863 & 25.35936 & 0.7236000 & 23.08319084 & 16.70299689 & 0.776100505 & 16.70299689 \\
IonMonger 2.0 & 0.86032668 & 26.01898 & 0.7366500 & 24.14116573 & 17.78358974 & 0.794448540 & 17.78358974 \\
COMSOL & 0.84866273 & 25.35893 & 0.7236000 & 23.08278068 & 16.70270010 & 0.776105262 & 16.70270010 \\
OghmaNano & 0.84715122 & 25.41794 & 0.7200000 & 23.29384000 & 16.77156480 & 0.778883119 & 16.77156480 \\
\end{tabular}
\end{ruledtabular}
\end{table*}

\begin{figure*}[t]
\centering
\begin{minipage}{0.48\textwidth}
\centering
\includegraphics[width=\linewidth]{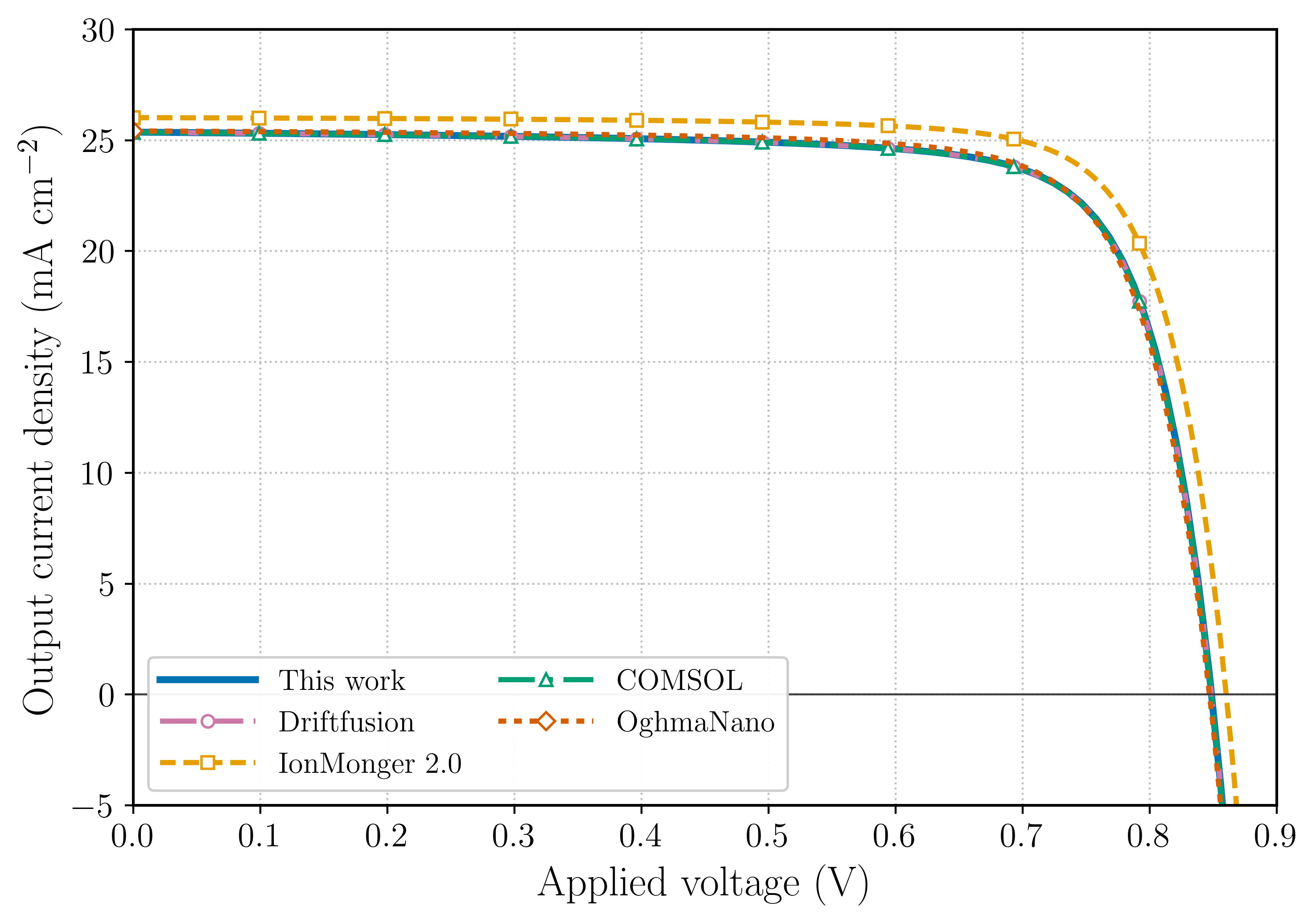}
\end{minipage}\hfill
\begin{minipage}{0.48\textwidth}
\centering
\includegraphics[width=\linewidth]{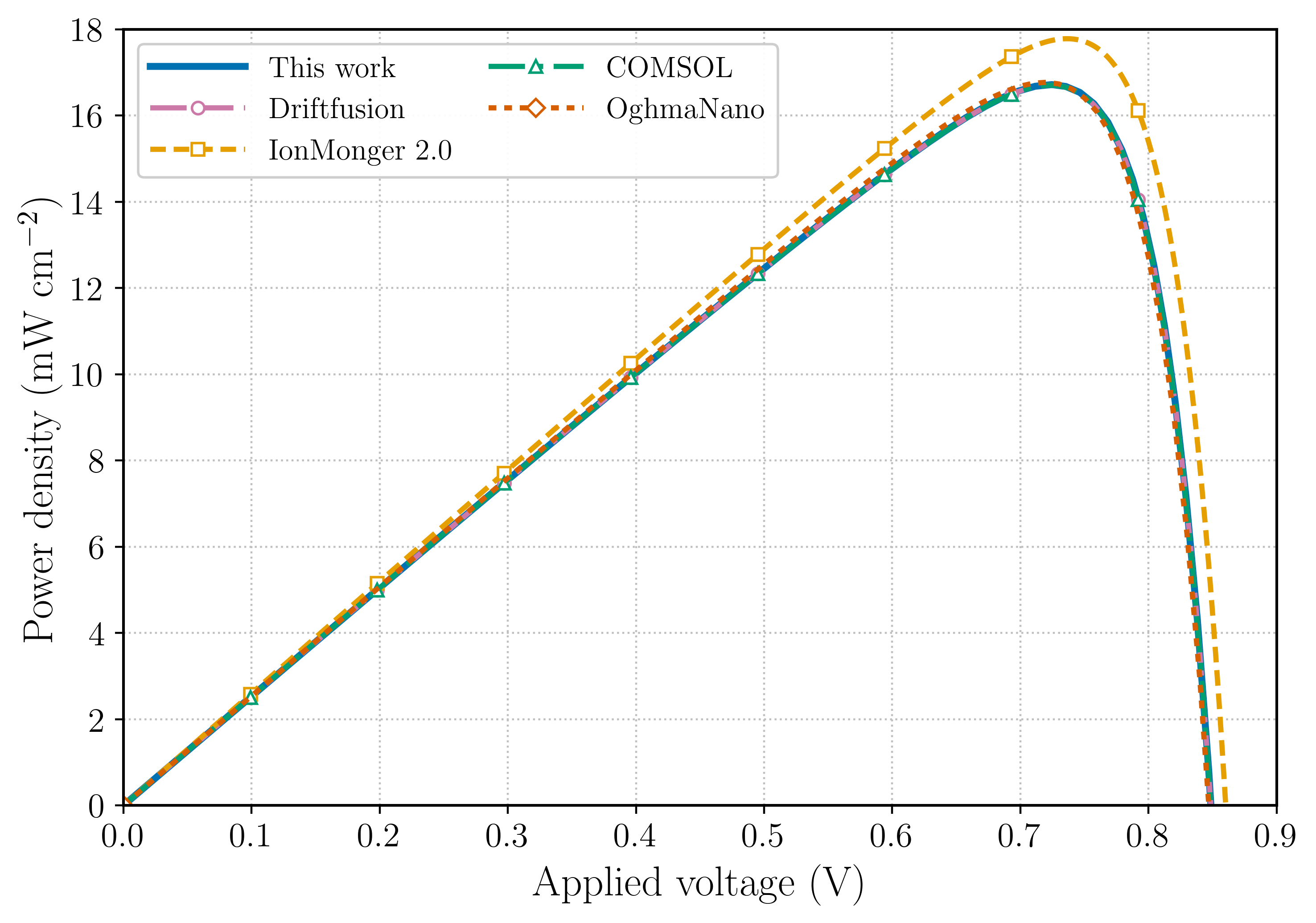}
\end{minipage}
\caption[Five-method device comparison]{Comparison of the $J$--$V$ characteristics and power-density curves from the five methods for the PNDIT-F3N/D18:L8-BO/PEDOT:PSS device over the common voltage range of 0--0.9~V.}
\label{compare-five-way}
\end{figure*}

Figure~\ref{compare-five-way}(a) and (b) show close agreement among this
work, Driftfusion, COMSOL, and OghmaNano over the operating range; IonMonger 2.0
is also consistent in the curve shape but gives a higher extracted power for this
parameter set. The values in Table~\ref{table-metrics} make the differences
explicit: the three closely agreeing reference curves have $P_{\max}$ between
16.703 and 16.772~mW~cm$^{-2}$, while IonMonger 2.0 gives 17.784~mW~cm$^{-2}$.
The five-way comparison supplies both the requested open-source and commercial
references under one device and voltage protocol. It is a reproducibility check
for this benchmark and does not imply a universal accuracy or speed ranking.

\subsection{Current conservation}

It is important to distinguish between numerical flux conservation and physical
steady-state current conservation, while recognizing how the former guarantees
the latter. In our finite-volume formulation, the discrete balance equations
provide a rigorous interpretation of conservation. For each control volume, the
change in stored electron or hole charge is exactly balanced by the net current
across the cell faces plus the integrated generation and recombination ($G-R$)
source terms. Because adjacent control volumes share the identical
Scharfetter--Gummel face flux, internal fluxes cancel exactly when the cell
equations are summed. Thus, charge is strictly conserved locally, limited only
by the specified nonlinear and linear solver tolerances.

In steady state, the electron and hole continuity equations reduce to
$\partial_x J_n=q(G-R)$ and $\partial_x J_p=-q(G-R)$. Summing these equations
cancels the source terms, dictating that the total current
$J=J_n+J_p$ must remain spatially constant, even as recombination continuously
redistributes the current between the two carrier components.

\begin{figure*}[t]
\centering
\includegraphics[width=0.98\textwidth]{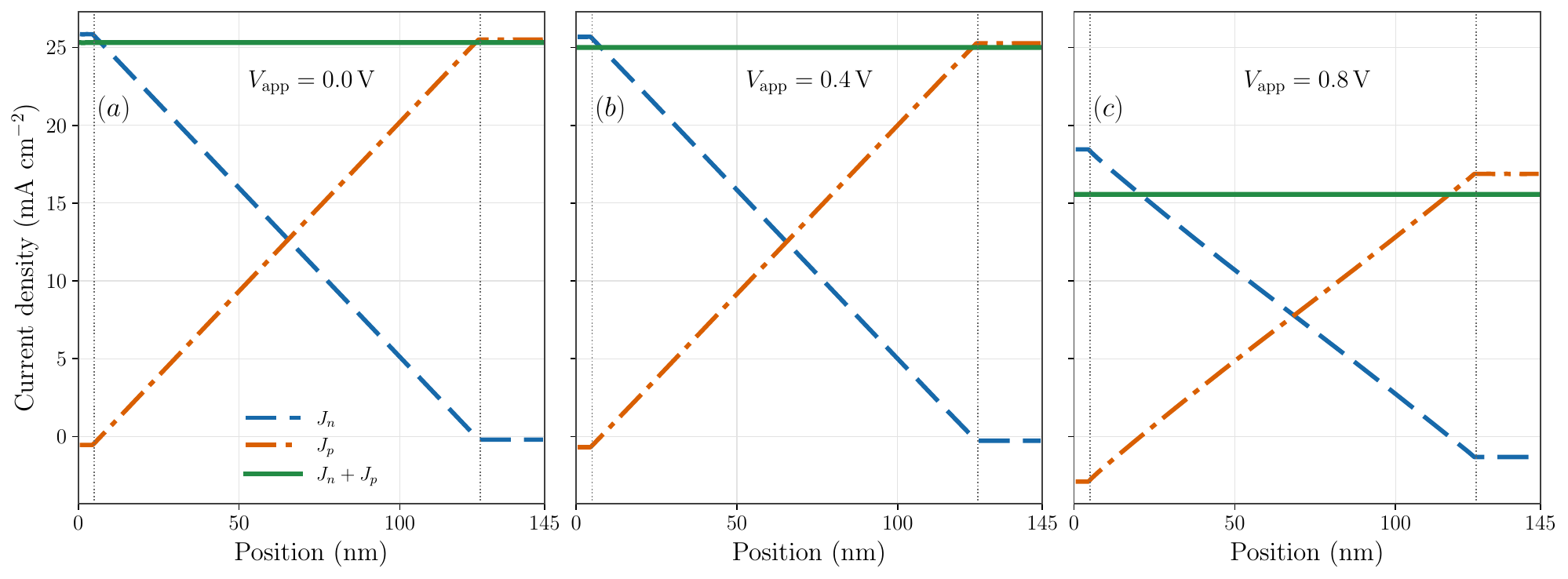}
\caption[Steady-state current conservation]{Steady-state electron ($J_n$), hole ($J_p$), and total ($J_n+J_p$)
current-density profiles as a function of position at applied biases of 0, 0.4,
and 0.8~V. The strict spatial uniformity of the total current at each bias
demonstrates rigorous steady-state current conservation, even as the individual
carrier components vary spatially due to recombination across the device.}
\label{current-conservation}
\end{figure*}

To verify this numerically, we simulated the multilayer device under applied
biases ranging from 0 to 1.0~V in 0.1~V increments, utilizing a fine 1~nm mesh
and a bias-stepping continuation method. Figure~\ref{current-conservation}
illustrates $J_n$, $J_p$, and the total current $J$ as functions of position for
representative biases. As theoretically expected, while the individual carrier
currents vary spatially due to recombination, the total current remains uniform.
The maximum relative spatial variation of the total current is strictly below
$5.7\times10^{-4}$ across all positions and simulated biases. Table~\ref{tab:current-conservation}
reports representative total current values at 0, 0.4, and 0.8~V. The spatial
variation of the individual carrier components, paired with the mathematically
constant total-current profile, rigorously demonstrates current conservation
within the discrete steady-state solution.

\begin{table}[t]
\caption{Steady-state current-conservation check for representative applied
biases. The mean total current is reported in mA cm$^{-2}$; the final column is
the maximum relative spatial variation of $J_n+J_p$.}
\label{tab:current-conservation}
\begin{ruledtabular}
\begin{tabular}{ccc}
$V_{\mathrm{app}}$ (V) & $\langle J_n+J_p\rangle$ (mA cm$^{-2}$) & $\max |\delta J|/|J|$ \\
\colrule
0.0 & 25.310 & $4.50\times10^{-4}$ \\
0.4 & 24.999 & $4.62\times10^{-4}$ \\
0.8 & 15.559 & $5.65\times10^{-4}$
\end{tabular}
\end{ruledtabular}
\end{table}

\subsection{Exciton dynamics}

To demonstrate the capability of the numerical framework in handling complex
kinetic processes, we implemented a zero-dimensional (0D) exciton dynamics
model {\cite{azzouzi2022reconciling}} to simulate the transient evolution of
localized exciton (LE) and charge-transfer (CT) states as well as charge
separated (CS) states in organic solar cells. The model is based on the rate
equations described by Eqs.(\ref{le})--(\ref{ct}), depicting the generation,
transfer, recombination, and dissociation processes of LE, CT and CS states
through  coupling with drift-diffusion-poisson equations.

\begin{table*}[t]
\caption{Key parameters for the 0-D exciton kinetic model coupled with drift-diffusion-poisson system.}
\label{table-exciton}
\begin{ruledtabular}
\scriptsize
\begin{tabular}{llcccc}
Quantity & Symbol & ETL & Active Layer & HTL & Unit \\
\colrule
Layer thickness & $d$ & 100 & 100 & 100 & nm \\
Relative permittivity & $\varepsilon_r$ & 3.0 & 3.0 & 3.0 & -- \\
Electron diffusion coefficient & $D_n$ & $7.7\times10^{-5}$ & $7.7\times10^{-6}$ & $7.7\times10^{-9}$ & $\mathrm{m^2\,s^{-1}}$ \\
Hole diffusion coefficient & $D_p$ & $7.7\times10^{-9}$ & $7.7\times10^{-6}$ & $7.7\times10^{-5}$ & $\mathrm{m^2\,s^{-1}}$ \\
Conduction-band eDOS & $N_c$ & $2\times10^{25}$ & $2\times10^{25}$ & $2\times10^{25}$ & $\mathrm{m^{-3}}$ \\
Valence-band eDOS & $N_v$ & $2\times10^{25}$ & $2\times10^{25}$ & $2\times10^{25}$ & $\mathrm{m^{-3}}$ \\
Effective density of states for LE & $g_{\tmop{LE}}$ & -- & $8\times10^{25}$ & -- & $\mathrm{m^{-3}}$ \\
Effective density of states for CT & $g_{\tmop{CT}}$ & -- & $4\times10^{25}$ & -- & $\mathrm{m^{-3}}$ \\
Free energy difference for CT & $\Delta G_{\tmop{CT}}$ & -- & 1.31 & -- & eV \\
Free energy difference between LE and CT state & $\Delta G_{\tmop{LE},\tmop{CT}}$ & -- & 0.32 & -- & eV \\
Free energy difference between CT and CS state & $\Delta G_{\tmop{CT},\tmop{CS}}$ & -- & 0.12 & -- & eV \\
Electron affinity & $\Phi_{\tmop{EA}}$ & 0.0 & 0.0 & 0.0 & eV \\
Ionisation potential & $\Phi_{\tmop{IP}}$ & $-1.19$ & $-1.19$ & $-1.19$ & eV \\
Equilibrium Fermi level & $E_{F0}$ & $-0.1$ & $-0.595$ & $-1.09$ & eV \\
LE$\rightarrow$CT transfer rate & $k_{\tmop{trans}}^{\tmop{LE}\rightarrow\tmop{CT}}$ & -- & $1.7\times10^{11}$ & -- & $\mathrm{s^{-1}}$ \\
CT$\rightarrow$LE transfer rate & $k_{\tmop{trans}}^{\tmop{CT}\rightarrow\tmop{LE}}$ & -- & $1\times10^9$ & -- & $\mathrm{s^{-1}}$ \\
LE recombination rate & $k_{\tmop{rec}}^{\tmop{LE}}$ & -- & $5.39\times10^9$ & -- & $\mathrm{s^{-1}}$ \\
CT recombination rate & $k_{\tmop{rec}}^{\tmop{CT}}$ & -- & $7\times10^9$ & -- & $\mathrm{s^{-1}}$ \\
CT dissociation rate & $k_{\tmop{diss}}^{\tmop{CT}\rightarrow\tmop{CS}}$ & -- & $2\times10^{10}$ & -- & $\mathrm{s^{-1}}$ \\
Free charge to CT state rate constant & $B_{\tmrsub{for}}^{\tmop{CS}\rightarrow\tmop{CT}}$ & -- & $7.4\times10^{-16}$ & -- & $\mathrm{m^3\,s^{-1}}$ \\
Equilibrium LE concentration & $[\tmop{LE}]_0$ & -- & $3.3\times10^{-2}$ & -- & $\mathrm{m^{-3}}$ \\
Equilibrium CT concentration & $[\tmop{CT}]_0$ & -- & $5.6\times10^{1}$ & -- & $\mathrm{m^{-3}}$ \\
Laser generation rate & $G$ & 0 & $1.1\times10^{29}$ & 0 & $\mathrm{m^{-3}\,s^{-1}}$ \\
Left boundary electron recomb. velocity & $v_{n,l}$ & $1.0\times10^{5}$ & -- & -- & $\mathrm{m^2\,s^{-1}}$ \\
Left boundary hole recomb. velocity & $v_{p,l}$ & 10 & -- & -- & $\mathrm{m^2\,s^{-1}}$ \\
Right boundary electron recomb. velocity & $v_{n,r}$ & -- & -- & 10 & $\mathrm{m^2\,s^{-1}}$ \\
Right boundary hole recomb. velocity & $v_{p,r}$ & -- & -- & $1.0\times10^{5}$ & $\mathrm{m^2\,s^{-1}}$
\end{tabular}
\end{ruledtabular}
\end{table*}

\begin{figure}[b]
\includegraphics[width=\columnwidth]{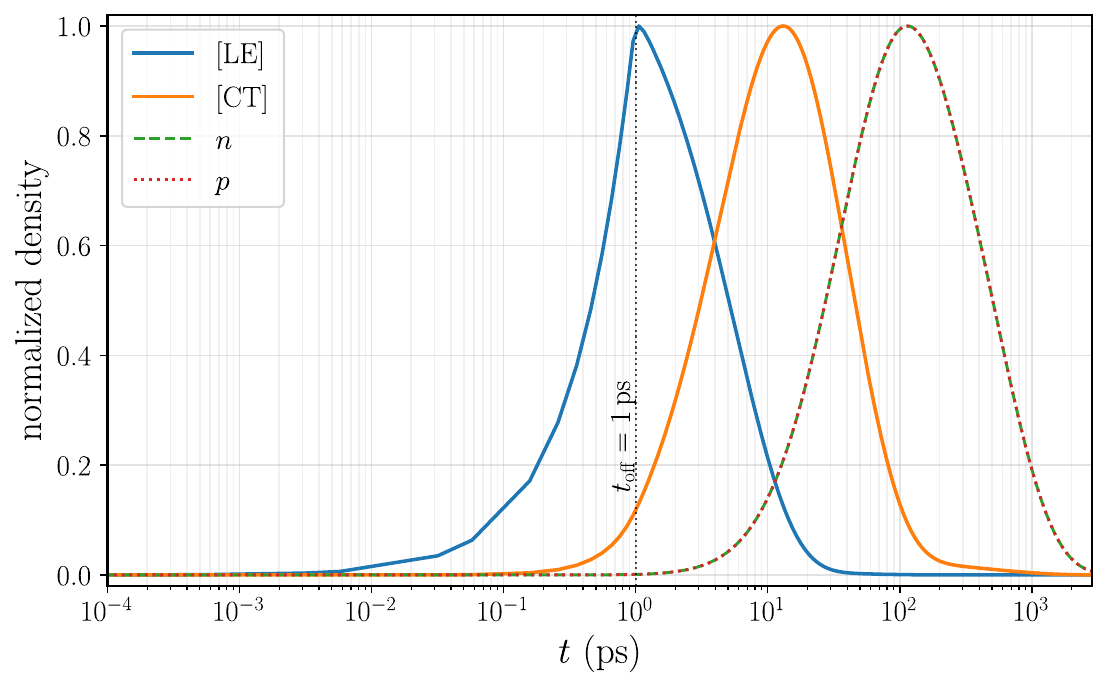}
\caption{Transient response for normalized $[\tmop{LE}]$, $[\tmop{CT}]$, and free carriers at the active-layer midpoint under a 0--1 ps light pulse.}
\label{exciton-pulse}
\end{figure}

To evaluate the decay kinetics of the different species, we apply a 1 ps
square light pulse (0--1 ps) under open-circuit conditions. All kinetic
parameters are taken from Table \ref{table-exciton} and are constrained to
satisfy detailed balance, ensuring a thermodynamically self-consistent
parameter set. Figure \ref{exciton-pulse} presents the transient responses of
the normalized $[\tmop{LE}]$, $[\tmop{CT}]$, and the free-carrier densities
$n$ and $p$ at the midpoint of the active layer, plotted versus time on a
logarithmic axis. Upon illumination, $[\tmop{LE}]$ rises first, indicating
that photogeneration initially populates local excitons. After the pulse is
switched off at $t_{\tmop{off}} = 1$ps, $[\tmop{LE}]$ decays while
$[\tmop{CT}]$ continues to rise, indicating ongoing LE$\rightarrow$CT
conversion. At later times, $[\tmop{CT}]$ also starts to decay, whereas the
free-carrier densities $n$ and $p$ build up and subsequently decay due to
recombination. The sequential rise and decay of LE, CT, and CS population
indicates the dynamic interaction between those states following the state
diagram in Figure \ref{sd}. The results demonstrate that the proposed
numerical framework can capture exciton dynamics relevant to excitonic
photovoltaics, particularly organic solar cells.

\subsection{Transient current decay}

Referring to {\cite{courtier2018fast}}, we designed a voltage scan protocol as
\begin{equation}
  V_{\tmop{app}} = V_{\tmop{bi}}  \left[ 1 - \frac{\tanh (\chi t)}{\tanh (\chi
  t_f)} \right], \qquad \tmop{with} \chi = 0.1, t_f = 10^4,
\end{equation}
and assume monomolecular recombination (depending solely on the local hole
concentration $p$) alone
\begin{equation}
  R (n, p) = \gamma p,
\end{equation}
taking $\gamma = 1500$ (same magnitude as reference) to simulate a transient
current behavior. And the total current calculated by
\begin{equation}
  \tmmathbf{J}=\tmmathbf{J}_n +\tmmathbf{J}_p +\tmmathbf{J}_P +\tmmathbf{J}_d,
\end{equation}
where $\tmmathbf{J}_d = \frac{\partial E}{\partial t}$ is displacement
current.

\begin{figure}[t]
\includegraphics[width=\columnwidth]{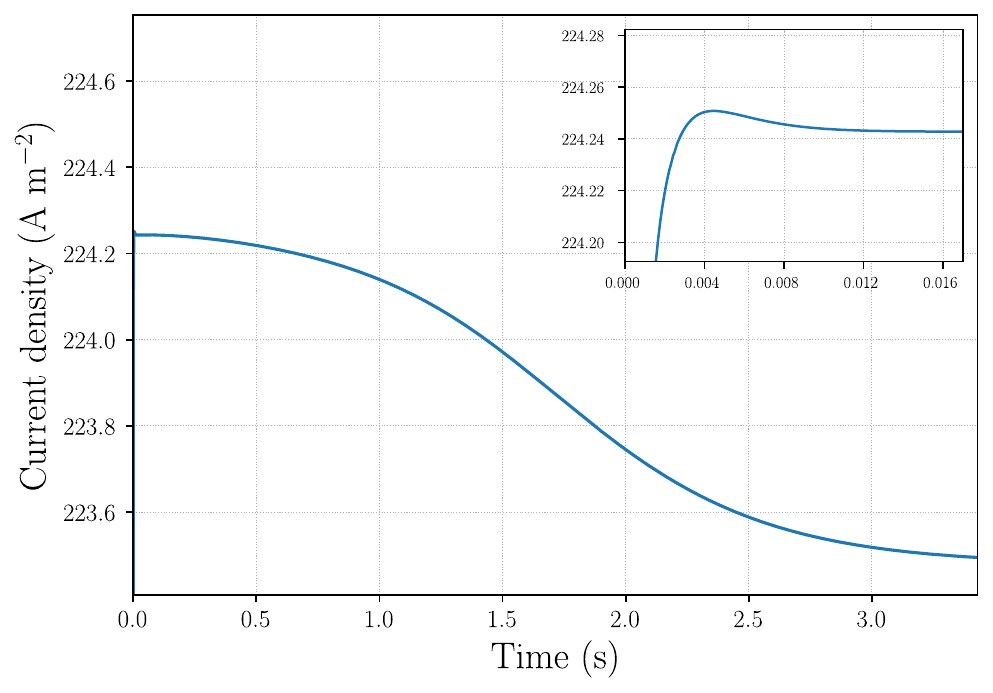}
\caption{Current density versus time at left boundary response under a fast decay voltage scan protocol.}
\label{transientc}
\end{figure}

Figure \ref{transientc} shows the total current density $\tmmathbf{J} (t)$ as
a function of the dimensional time $t$ (in seconds). The main panel
illustrates the slow relaxation of the current towards its steady value over a
macroscopic time interval of a few seconds, while the inset zooms into the
initial short-time window and reveals a rapid overshoot followed by a monotone
decay. Although the time axis used here differs from that in
{\cite{courtier2018fast}}, the overall transient behaviour closely resembles
their Figure 7. In particular, our computation clearly resolves the fast
electronic transient that is not present in the asymptotic approximation of
{\cite{courtier2019systematic}}, demonstrating that the proposed
time-integration scheme can accurately capture such short-time-scale dynamics.

\subsection{\texorpdfstring{$J$--$V$ hysteresis in perovskite solar cell with mobile ions}{J-V hysteresis in perovskite solar cell with mobile ions}}

\begin{table*}[!t]
\caption{Key model parameters for the three-layer perovskite solar cell.}
\label{psc}
\begin{ruledtabular}
\scriptsize
\begin{tabular}{llcccc}
Quantity & Symbol & ETL & Perovskite & HTL & Unit \\
\colrule
Layer thickness & $d$ & 100 & 600 & 200 & nm \\
Relative permittivity & $\varepsilon_r$ & 10.0 & 24.1 & 3.0 & -- \\
Electron diffusion coefficient & $D_n$ & $1.0\times10^{-5}$ & $1.7\times10^{-4}$ & $1.0\times10^{-6}$ & $\mathrm{m^2\,s^{-1}}$ \\
Hole diffusion coefficient & $D_p$ & $1.0\times10^{-5}$ & $1.7\times10^{-4}$ & $1.0\times10^{-6}$ & $\mathrm{m^2\,s^{-1}}$ \\
Conduction-band DOS & $N_c$ & $5.0\times10^{25}$ & $8.1\times10^{24}$ & $5.0\times10^{25}$ & $\mathrm{m^{-3}}$ \\
Valence-band DOS & $N_v$ & $5.0\times10^{25}$ & $5.8\times10^{24}$ & $5.0\times10^{25}$ & $\mathrm{m^{-3}}$ \\
Electron affinity & $\Phi_{\tmop{EA}}$ & $-4.0$ & $-3.7$ & $-3.1$ & eV \\
Ionisation potential & $\Phi_{\tmop{IP}}$ & $-6.0$ & $-5.4$ & $-5.1$ & eV \\
Equilibrium Fermi level & $E_{F0}$ & $-4.1$ & $-4.55$ & $-5.0$ & eV \\
Trap energy level & $E_{\tmop{trap}}$ & $-4.25$ & $-4.55$ & $-4.75$ & eV \\
SRH lifetime (electrons) & $\tau_n$ & -- & $3.0\times10^{-9}$ & -- & s \\
SRH lifetime (holes) & $\tau_p$ & -- & $3.0\times10^{-9}$ & -- & s \\
Bulk bimolecular recomb. & $\beta$ & 0 & $1.0\times10^{-18}$ & 0 & $\mathrm{m^3\,s^{-1}}$ \\
Ionic vacancy density & $N_I$ & -- & $1.6\times10^{25}$ & -- & $\mathrm{m^{-3}}$ \\
Ionic diffusion coefficient & $D_I$ & 0 & $1.0\times10^{-17}$ & 0 & $\mathrm{m^2\,s^{-1}}$ \\
Uniform generation rate & $G$ & 0 & $2.5\times10^{27}$ & 0 & $\mathrm{m^{-3}\,s^{-1}}$ \\
Left boundary electron recomb. velocity & $v_{n,l}$ & $1.0\times10^{5}$ & -- & -- & $\mathrm{m^2\,s^{-1}}$ \\
Left boundary hole recomb. velocity & $v_{p,l}$ & 10 & -- & -- & $\mathrm{m^2\,s^{-1}}$ \\
Right boundary electron recomb. velocity & $v_{n,r}$ & -- & -- & 10 & $\mathrm{m^2\,s^{-1}}$ \\
Right boundary hole recomb. velocity & $v_{p,r}$ & -- & -- & $1.0\times10^{5}$ & $\mathrm{m^2\,s^{-1}}$
\end{tabular}
\end{ruledtabular}
\end{table*}

Using the parameter set in Table \ref{psc}, we simulate a three-layer
perovskite solar cell with mobile ionic vacancies in the absorber. The optical
generation is treated as uniform in the perovskite layer with a volume
generation rate $G = 2.5 \times 10^{27} \text{ } m^{- 3} \text{ } s^{- 1}$,
which yields a short-circuit current density $J_{\tmop{sc}} \approx 22
\text{~} \tmop{mA} \text{ } \tmop{cm}^{- 2}$, comparable to experimentally
reported values for high-performance perovskite devices. Ionic vacancies with
density $N_0 = 1.6 \times 10^{25} \text{ } m^{- 3}$ and diffusion coefficient
$D_I = 10^{- 17} \text{ } m^2 \text{ } s^{- 1}$ are included only in the
perovskite layer, while the transport layers are assumed ion-immobile.

\begin{figure}[t]
\includegraphics[width=\columnwidth]{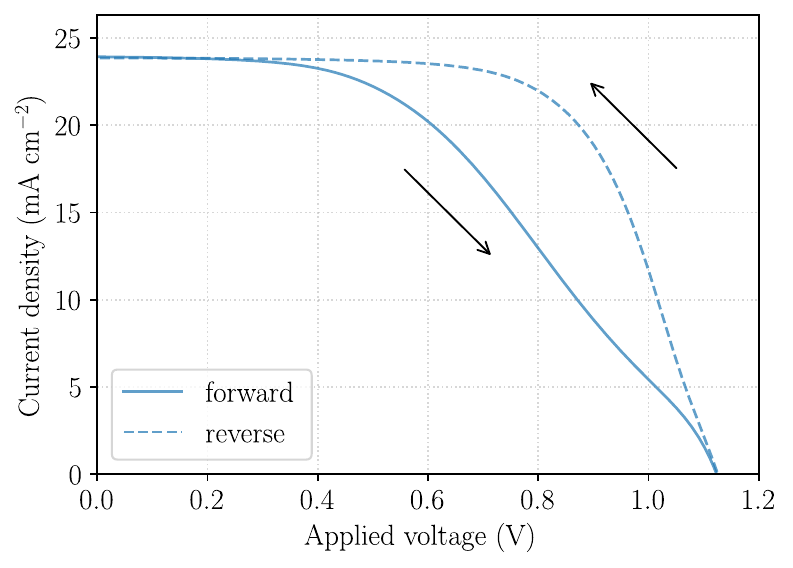}
\caption{Simulated current--voltage characteristics of a three-layer perovskite solar cell with mobile ions at a scan rate of 0.1 $V\,s^{-1}$. The device consists of an electron-transport layer (ETL, 100 nm), a perovskite absorber (600 nm) and a hole-transport layer (HTL, 200 nm). Mobile ionic vacancies are included only in the perovskite layer. Solid and dashed lines denote forward (0 $\rightarrow$ 1.2~V) and reverse (1.2 $\rightarrow$ 0~V) scans, respectively; the black arrows indicate the corresponding scan directions. A clear $J$--$V$ hysteresis loop is obtained within the same drift--diffusion--Poisson framework.}
\label{pscfig}
\end{figure}

Figure \ref{pscfig} shows the resulting $J$--$V$ characteristics under a
voltage scan from 0 to 1.2 V and back at a scan rate of 0.1 $V s^{- 1}$. The
forward and reverse sweeps display a pronounced hysteresis loop: during the
forward scan the current density decreases smoothly from $J_{\tmop{sc}}$ to
the vicinity of the open-circuit voltage, whereas the reverse scan exhibits a
much steeper rise in current and a delayed recovery near $V \approx 1 V$. This
behavior is entirely captured by the coupled drift--diffusion--Poisson--ionic
transport model without any additional empirical hysteresis terms.

The hysteresis originates from the slow redistribution of mobile ions, which
gradually screens the internal electric field and modifies the band bending at
the ETL/perovskite and perovskite/HTL interfaces. Under forward bias, ions
move in a direction that partially counteracts the built-in field, reducing
charge extraction and leading to lower current. When the scan direction is
reversed, the ionic configuration lags behind the applied voltage, so that the
internal field is temporarily closer to its initial state, resulting in a
higher current for the same applied voltage.

\section{Conclusion}

We have developed and validated a one-dimensional transient drift-diffusion
solver for multilayer solar cells that combines a finite-volume discretization
with Scharfetter--Gummel-type fluxes and a fifth-order Radau IIA implicit
Runge--Kutta scheme for time integration. The spatial discretization enforces
local charge conservation on general non-uniform meshes, handles sharp
heterojunctions and strongly varying material parameters in a natural way, and
is compatible with both density-based and potential-based formulations of the
van Roosbroeck system. The temporal discretization, implemented via a modified
Radau solver for index-1 DAEs, provides high-order accuracy and robust
L-stable damping of fast modes, which is crucial for resolving electronic and
ionic dynamics over many orders of magnitude in time.

Systematic convergence studies confirm that the proposed method attains
second-order accuracy in space and fifth-order accuracy in time for the
coupled Poisson and carrier-continuity equations. In an abrupt $p$--$n$
junction test, the numerical solution reproduces the classical depletion
approximation, including the built-in voltage and depletion widths, within
numerical precision. A transient test based on a fast voltage-decay protocol
shows that the scheme accurately captures both the rapid initial electronic
transient and the subsequent slow relaxation, in quantitative and qualitative
agreement with previously reported asymptotic and numerical results.
Furthermore, benchmarking against the existing software references OghmaNano
and COMSOL for a multilayer organic solar cell shows close agreement in the steady-state
$J$--$V$ response and the extracted photovoltaic metrics. The remaining
differences are small (below 1\% RMS at the curve level), confirming the
accuracy and reliability of our implementation for complex device structure.

To demonstrate both the extensibility of the framework and its physical
consistency, we include additional dynamic processes that are important in
emerging photovoltaic technologies. In organic solar cells, LE/CT state
kinetics are coupled to the drift--diffusion transport equations through local
generation and recombination terms, following Azzouzi et al.
{\cite{azzouzi2022reconciling}}. This coupling allows the model to resolve
exciton dynamics across multiple time scales, from picoseconds to nanoseconds.
In perovskite solar cells, mobile ionic vacancies in the absorber are treated
with a drift--diffusion equation driven by the Poisson electric field, which
naturally produces {\tmem{J}}--{\tmem{V}} hysteresis under voltage scans
without introducing empirical hysteresis terms and thereby reproduces a key
experimental signature. In both extensions, the added equations are
nondimensionalized and discretized in the same manner as the core
drift--diffusion system, and time integration is carried out with the same
high-order Radau IIA scheme.

Compared with existing DD simulators such as IonMonger, Driftfusion,
SolarDesign, and SIMsalabim, the present work emphasizes the combination of a
structure-preserving finite-volume discretization, high-order stiff time
integration, and algorithmic transparency. The parameter-convergence study also
shows robust steady-state behavior under mesh refinement and changes in applied
voltage, temperature, and contact doping. This behavior stems from the
combination of a locally conservative finite-volume formulation,
Scharfetter--Gummel flux discretization, damped Newton iteration, and L-stable
Radau IIA time integration. These features make the proposed method well suited
for efficient and reliable simulations of multilayer optoelectronic devices.

Overall, the numerical method developed here achieves a high temporal convergence
order, good numerical stability, and excellent computational efficiency. Its
capability has been extended to include excitons and mobile ions, enabling
device-scale simulations of modern solar cells. Although this work focuses
primarily on numerical method development, several worthwhile extensions are
readily apparent. On the numerical side, future efforts could explore tailored
preconditioners and block-structured linear solvers for the coupled stage systems,
as well as adaptive strategies that systematically balance spatial, temporal,
and algebraic errors. On the modeling side, the current one-dimensional framework
could be coupled with more detailed optical models, interface-specific
recombination schemes, and degradation mechanisms, or alternatively generalized
to two-dimensional device geometries. Transient simulation toolkits---such as
those for transient photovoltage, transient photocurrent, and transient
absorption measurements---could also be developed to facilitate the
interpretation of optoelectronic experimental results. Collectively, these
possibilities underscore the remarkable versatility of the numerical method
reported here.

\noindent\textit{Acknowledgements.} J.Y. acknowledges funding support from National Natural Science Foundation of China (No. 62574175 and No. 62404191), Guangdong Provincial Program (No. 2023QN10C144), Shenzhen Science and Technology Program (No. KQTD20240729102028011 and No. JCYJ20240813113553067), and Guangdong Basic Research Center of Excellence for Aggregate Science.

\textit{Data availability.} The data are available from the authors upon reasonable request.

\renewcommand{\bibfont}{\footnotesize}
\vspace{-\baselineskip}
\makeatletter
\let\write@bibliographystyle\relax
\makeatother

\end{document}